\documentclass[12pt]{article}
\usepackage{jheppub}
\pdfoutput = 1

\usepackage{amsmath}
\usepackage{amssymb}
\usepackage{bbm}
\usepackage{bbold}
\usepackage{braket}
\usepackage{caption}
\usepackage{color}
    \definecolor{darkgreen}{rgb}{0,0.5,0}
    \definecolor{darkred}{rgb}{0.5,0,0}
    \definecolor{darkblue}{rgb}{0,0,0.6}
    \definecolor{purple}{rgb}{0.4,.2,0.7}

\usepackage{float}
\usepackage{graphicx}
\usepackage{hyperref}

\usepackage{mathabx}
\usepackage{mathtools}
\usepackage{pdfsync}
\usepackage{slashed}
\usepackage[normalem]{ulem}
\usepackage{upgreek}
\usepackage{url}

\usepackage[ragged]{footmisc}
\def\be{\begin{equation}}
\def\ee{\end{equation}}

\renewcommand{\tilde}{\widetilde}

\numberwithin{equation}{section}

\begin{document}
%\subheader{empty}
\title{Inside an Asymptotically Flat Hairy Black Hole}
\author[a]{\'Oscar~J.C.~Dias,}
\author[b]{Gary~T.~Horowitz,}
\author[c]{Jorge~E.~Santos}
%affilations
\affiliation[a]{STAG research centre and Mathematical Sciences, University of Southampton, UK}
\affiliation[b]{Department of Physics, University of California at Santa Barbara, Santa Barbara, CA 93106, U.S.A.}
\affiliation[c]{Department of Applied Mathematics and Theoretical Physics, University of Cambridge, Wilberforce Road, Cambridge, CB3 0WA, UK}

% e-mail addresses
\emailAdd{O.J.Campos-Dias@soton.ac.uk}
\emailAdd{horowitz@ucsb.edu}
\emailAdd{jss55@cam.ac.uk}

\abstract{We study the interior of a recently constructed  family of asymptotically flat, charged black holes that develop (charged) scalar hair as one increases their charge  at fixed mass.   Inside the horizon, these black holes resemble the interior of a holographic superconductor. There are analogs of the Josephson oscillations of the scalar field, and the final Kasner singularity depends very sensitively on the black hole parameters near the onset of the instability. In an Appendix, we give a general argument that Cauchy horizons cannot exist in a large class of stationary black holes with scalar hair.}

\maketitle
%%%%%%%%%%%%%%%%%%
\section{Introduction} 

Static black holes in Einstein-Maxwell theory are very simple, and given by the Reissner-Nordstr\"om (RN) solutions. It has recently been shown that under certain conditions these black holes can become unstable to forming scalar hair, i.e., static scalar fields outside the horizon. One class of examples involves theories with higher curvature terms, such as scalar couplings to the Gauss-Bonnet action \cite{Doneva:2017bvd,Silva:2017uqg,Antoniou:2017acq}. In these examples, the Maxwell field plays no role and even Schwarzschild black holes can become unstable to forming neutral scalar hair due to the coupling of the scalar to the curvature near the horizon. Another class of examples without higher curvature terms involves coupling a massless scalar field directly to $F_{ab}F^{ab}$ \cite{Herdeiro:2018wub,Fernandes:2019rez}. In this case, when the electric charge on the black hole becomes large enough, RN becomes unstable. This is because $F_{ab}F^{ab} < 0$ for an electrically charged black hole, and acts like a negative potential for the scalar near the horizon. There is growing literature on these ``scalarized" black holes (see references in \cite{LuisBlazquez-Salcedo:2020rqp} for a rather recent list of papers) but it almost always is focussed on neutral scalar fields.

In a recent paper \cite{Dias:2021vve}, we added a massive, charged scalar field $\psi$ to Einstein-Maxwell with a simple $|\psi|^2 F_{ab}F^{ab}$ coupling. As above, when the electric charge is large enough, RN becomes unstable and develops charged scalar hair. We found that the hairy black holes in this theory can have unusual extremal limits. Previously studied hairy black holes either become singular in the extremal limit, or have a zero temperature degenerate horizon like RN. In the theory with $|\psi|^2 F_{ab}F^{ab}$ coupling,  we found that for some range of parameters,  the solution with maximum charge is a nonsingular black hole with nonzero temperature. We called such objects ``maximal warm holes".

In this paper, we look inside the horizon of these hairy black 
holes. These solutions  are asymptotically flat analogs of the asymptotically anti-de Sitter (AdS) solutions known as holographic superconductors \cite{Gubser:2008px,Hartnoll:2008vx,Hartnoll:2008kx}, which have been extensively studied. In AdS, the charged scalar condenses at low temperature without any explicit coupling between the scalar and Maxwell field. Without the cosmological constant, however, this does not happen \cite{Hod:2015hza} and one needs to add an interaction like  $|\psi|^2 F_{ab}F^{ab}$.\footnote{Even without a direct coupling between the scalar and Maxwell field, there are black holes with charged scalar hair if the scalar is self-interacting \cite{Herdeiro:2020xmb,Hong:2020miv}. However those black holes do not branch off from RN, so the RN solution never becomes unstable. In higher dimensions, there are examples of hairy black holes in  Einstein-Gauss-Bonnet gravity with a minimally coupled charged scalar field with no self interactions \cite{Grandi:2017zgz}.}

It was recently shown \cite{Hartnoll:2020fhc} that there is interesting dynamics inside the horizon of a holographic superconductor. Just below the critical temperature $T_c$ when the scalar field is very small outside, the interior evolves through several distinct epochs, including a collapse of the Einstein-Rosen bridge, and Josephson oscillations of the scalar field. There is no Cauchy horizon and the solution ends in a Kasner singularity, which is characterized by one parameter $p_t$. Most importantly, it was shown that $p_t$ is extremely sensitive to the black hole temperature near $T_c$: for any $\epsilon > 0 $, $p_t$ cycles through a finite range of values an infinite number of times as $T$ increases from $T_c -\epsilon $ to $T_c$.

To see whether this sensitive dependence of the black hole singularity on its temperature is just a feature of AdS black holes, we study the interior of the solutions constructed in \cite{Dias:2021vve}.\footnote{For other recent discussions of the interior of hairy black holes, see \cite{Grandi:2021ajl,Mansoori:2021wxf}.}  We find almost exactly the same behavior as seen in the holographic superconductors. In particular, near the onset of the instability, one again sees the same dynamical epochs and extreme sensitivity of the singularity on the black hole parameters. The main difference between the AdS and asymptotically flat black holes is in the transition from one Kasner regime to another that can occur inside the horizon. In our asymptotically case, these depend on the new coupling between the scalar and Maxwell field, and are more complicated than for AdS black holes.

The fact that most hairy black holes do not have Cauchy horizons has been established in a series of papers \cite{Hartnoll:2020rwq,Hartnoll:2020fhc,Cai:2020wrp,Devecioglu:2021xug,An:2021plu,VandeMoortel:2021gsp}. We give a proof applicable to our theory in Sec. 4. In addition,  we give a more general argument in an Appendix which applies to all stationary solutions of any theory of gravity coupled to a complex scalar with positive potential. In particular, it applies to the rotating black holes with scalar hair constructed in \cite{Herdeiro:2014goa,Herdeiro:2015gia,Chodosh:2015nma,Chodosh:2015oma}. To our knowledge this is the first argument that applies to solutions that are not static and either spherically symmetric or planar, \emph{i.e.}, with cohomogeneity greater than one.
%%%%%%%%%%%%%%%%%%

%%%%%%%%%%%%%%%%%%
\section{Equations of motion}
%%%%%%%%%%%%%%%%%%

We are interested in asymptotically flat charged black hole solutions that can develop scalar hair due to a particular coupling of a massive charged scalar field to the Maxwell field. So we consider the action  
\begin{equation}\label{eq:action}
S= \int \mathrm{d}^{4}x \sqrt{-g}\left[R- F^2-4(\mathcal{D}_a\psi)(\mathcal{D}^a \psi)^\dagger-4 m^2 |\psi|^2-4 \alpha F^2 |\psi|^2\right]\,,
\end{equation}
where $m$ and $q$ are the mass and charge of the scalar field and $\mathcal{D}=\nabla-i\,q\,A$ and $F=\mathrm{d}A$\,. This theory satisfies all the usual energy conditions  if the Maxwell-scalar coupling constant $\alpha$ is positive, which we will assume is the case.

The equations of motion for this action read
\begin{subequations}\label{EOM:S}
\begin{multline}
R_{ab}-\frac{R}{2}g_{ab}=2\left(1+4 \alpha |\psi|^2\right)\left(F_{ac}F_b^{\phantom{b}c}-\frac{g_{ab}}{4}F^{cd}F_{cd}\right)
\\
+2\left[(\mathcal{D}_a \psi) (\mathcal{D}_b \psi)^\dagger+(\mathcal{D}_a \psi)^\dagger(\mathcal{D}_b \psi) -g_{ab} (\mathcal{D}_c \psi)(\mathcal{D}^c \psi)^\dagger-g_{ab} m^2 |\psi|^2  \right]\,,
\end{multline}
\begin{equation}
\nabla_a\left[\left(1+4 \alpha |\psi|^2\right)F^{ab}\right]=i\,q\, \left[(\mathcal{D}^b \psi)\psi^\dagger-(\mathcal{D}^b \psi)^\dagger \psi \right]\,,
\end{equation}
\begin{equation}
\mathcal{D}_a \mathcal{D}^a \psi-\alpha F^{cd}F_{cd} \psi-m^2 \psi=0\,.
\end{equation}
\end{subequations}

Besides the Reissner-Nordstr\"om solution with vanishing scalar field, this theory also has hairy black hole solutions with nonzero $\psi$. In fact, as we increase the charge on a RN black hole, it becomes unstable at a critical charge $Q_c$. This instability and the resulting hairy black holes were studied  (outside the  horizon) in \cite{Dias:2021vve}. Here, we are interested in diving into the event horizon of these hairy black holes and studying the properties of their interior.

To study the interior of the asymptotically flat charged black holes of \eqref{eq:action}  it is convenient to use the ansatz\footnote{To make contact with the line element that we used in the companion paper  \cite{Dias:2021vve}, \eqref{ansatz} reduces to it when we redefine the radial coordinate $z=r_+/r$ and set $f=z^2 p$ and $\chi = -2 \ln g$. Moreover, setting $\kappa=1$, the change of variable $x=\cos \theta$ rewrites the 2-sphere line element in the familiar spherical coordinates.}
\begin{subequations}
\label{ansatz}
\begin{align}
& \mathrm{d}s^2 = \frac{r_+^2}{z^2}\left[-f(z) e^{-\chi(z)}\frac{\mathrm{d}t^2}{r_+^2}+\frac{\mathrm{d}z^2}{f(z)}+\frac{\mathrm{d}x^2}{1-\kappa\,x^2}+(1-\kappa\,x^2)\mathrm{d}\phi^2\right]
\\
& A = \,\Phi(z)\,\mathrm{d}t\,,
\\
& \psi=\psi^\dagger=\psi(z)\,,
\end{align}
\end{subequations}
where $\kappa=1$,\footnote{Introducing $\kappa$ will help us to later identify  terms coming from the curvature of the two-spheres in the equations of motion. Unlike in AdS space where solutions exist with $\kappa = \pm 1, 0$,  only spherical spatial cross sections yield solutions when  $\Lambda=0$.} and $f(z)$, $\chi(z)$, $\Phi(z)$ and $\psi(z)$ are function of $z$ only. The parameter $r_+$ controls the temperature of the event horizon and the area of this bifurcating Killing horizon, as displayed below in \eqref{Thermo}.

We will require that $f(1) = 0$, so non-extremal charged black hole solutions described by \eqref{ansatz} have an event horizon at $z=z_{\mathcal{H}}\equiv 1$. We will show in the next section that there is no Cauchy horizon when $\psi$ is nonzero.
%and a Cauchy horizon at $z=z_{\mathcal{I}}>1$.
The asymptotic region is at $z=0$ and thus the exterior of the black hole is the region $z\in(0,1)$ while the interior region between the event horizon and singularity  is  $z\in(1,\infty)$. The coordinate $z$ is a spacelike radial coordinate in the exterior region but it becomes timelike in the interior region.

Inserting \eqref{ansatz} into \eqref{EOM:S}, the equations of motion boil down to 
\begin{subequations}\label{eom}
\begin{align}
& z^2 e^{-\frac{\chi}{2}}\left[\left(1+4 \alpha  \psi^2\right) e^{\frac{\chi}{2}} \Phi^\prime\right]^\prime-\frac{2 \tilde{q}^2y_+^2 \psi^2 \Phi}{f}=0\,, \label{eomPhi}
\\
& z^2 e^{\frac{\chi}{2}} \left(\frac{e^{-\frac{\chi}{2}} f \psi^\prime}{z^2}\right)^\prime-\left(\frac{y_+^2}{z^2}-\frac{\tilde{q}^2 y_+^2 e^{\chi} \Phi^2}{f}-2 e^{\chi} z^2 \alpha  {\Phi^\prime}^2\right) \psi=0\,, \label{eomPsi}
\\
& \chi^\prime-4\,z\,\left(\frac{\tilde{q}^2y_+^2 e^{\chi}}{f^2} \psi ^2 \Phi ^2+{\psi^\prime}^2\right)=0\,, \label{eomChi}
\\
& e^{\frac{\chi}{2}} z^4 \left(\frac{e^{-\frac{\chi}{2}} f}{z^3}\right)^\prime+z^2 \kappa-2 y_+^2 \psi^2-\left(1+4 \alpha  \psi^2\right) e^{\chi} z^4 {\Phi^\prime}^2=0\,, \label{eomf}
\end{align}
\end{subequations}%%
where $\tilde{q} \equiv q/m$,  $y_+ \equiv r_+ m$, and ${}^\prime$ denotes a derivative with respect to $z$.
 The term proportional to $\kappa=1$ in the last  equation comes from the curvature of the two-spheres.

To solve numerically the equations of motion, we will find convenient to do a field redefinition that explicitly indicates that the solution has an event horizon at $z=1$ and we are working in a gauge where $A_t$ vanishes on the horizon (i.e. $q_1$ and $q_2$ are finite at $z=1$):
\begin{equation}
f(z)=z^2\left(1-z\right)q_1(z)\,,\quad \Phi(z) = \left(1-z\right)q_2(z)\,,\quad \psi(z)=q_3(z)\,, \quad \chi(z)=-\ln q_4(z)\,.
\end{equation}

Let $\mu \equiv A_t (0) = q_2(0)$ denote the electrostatic potential at infinity.
 The dimensionless mass $M$, charge $Q$, entropy $S_\mathcal{H}$ and temperature $T_\mathcal{H}$ of the black holes are then given by:
\begin{eqnarray} \label{Thermo}
&&  M m = \frac{y_+}{2}[1-q^\prime_1(0)]\,,
\qquad
Q m = y_+\,[\mu-q^\prime_2(0)]\,, \nonumber\\
&& S_\mathcal{H}\, m^2=\pi\,y_+^2\quad\text{and}\quad \frac{T_\mathcal{H}}{m}=\frac{q_1(1)\sqrt{q_4(1)}}{4\pi y_+}\,.
\end{eqnarray}

We work with dimensionless quantities since there is a scaling symmetry which relates different solutions. Inequivalent soutions can be labeled by the four dimensionless quantities $ \tilde q, \alpha, Mm,  Qm$. However, to find the solutions numerically, it is more convenient to use a slightly different set of dimensionless quantities: $(\tilde{q},\alpha, y_+, \mu)$.

When the scalar field vanishes, \emph{i.e.} $\psi=0$, the only charged  black hole of the theory is given by the familiar Reissner-Nordstr\"om (RN) solution with event horizon at $z=z_{\mathcal{H}}\equiv 1$ and Cauchy horizon at $z=z_{\mathcal{I}}=1/\mu^2$.
In our coordinates, this solution is: 
 \begin{equation}\label{RNsol:int}
f(z)=\frac{1}{z_{\mathcal{I}}} z^2(z-1)(z-z_{\mathcal{I}})\,,\quad \chi(z)=0\,,\quad\text{and}\quad \Phi(z)=\mu \left(1-z\right).
\end{equation}

%%%%%%%%%%%%%%%%%%%%%%%
\section{No smooth inner horizon for charged fields \label{sec:theorem}}

Before discussing the interior dynamics, we first show that these black holes cannot have a smooth inner horizon (similar proofs were first presented in \cite{Hartnoll:2020fhc,Cai:2020wrp}).
This follows from the existence of a conserved quantity in our theory. By virtue of the equations of motion \eqref{eom} the following quantity is a constant
\begin{equation}
C_1 = \frac{e^{\frac{\chi}{2}}}{z^2}\left(e^{-\chi}\, f\right)^\prime-4 e^{\frac{\chi}{2}}\Phi^\prime\,\Phi(1+4\,\alpha\,\psi^2)+2 \kappa \int_{z_{\mathcal{H}}}^{z}\frac{e^{-\frac{\chi(x)}{2}}}{x^2}\mathrm{d}x\,.
\end{equation}
Let us assume an inner horizon exists at $z=z_{\mathcal{I}}>z_{\mathcal{H}}$. Then, since $f(z)<0$ when $z\in(z_{\mathcal{H}},z_{\mathcal{I}})$ and must vanish at a horizon, we need to have $f^\prime(z_{\mathcal{I}})>0$ at $z=z_{\mathcal{I}}$. From the form of the equations of motion, it is clear that a smooth horizon must have either $\Phi = 0$ or $\psi = 0$. But if $\psi = 0$ on the horizon, it must vanish everywhere near the horizon (here we implicitly assume that $\psi $ is analytic in a neighborhood of the horizon). So we must have $\Phi(z_{\mathcal{I}})=0$.

Evaluating $C_1$ at the event horizon gives
\begin{equation}
C_1 = \frac{e^{\frac{\chi(z_{\mathcal{H}})}{2}}}{z_{\mathcal{H}}^2}f^\prime(z_{\mathcal{H}})<0
\end{equation}
where we used that $f^\prime(z_{\mathcal{H}})<0$ since $z=z_{\mathcal{H}}$ is a smooth black hole horizon. At the inner horizon, we find
\begin{equation}
C_1 = \frac{e^{\frac{\chi(z_{\mathcal{I}})}{2}}}{z_{\mathcal{I}}^2}f^\prime(z_{\mathcal{I}})+2 \kappa \int_{z_{\mathcal{H}}}^{z_{\mathcal{I}}}\frac{e^{-\frac{\chi(x)}{2}}}{x^2}\mathrm{d}x>0\,,
\end{equation}
but since $C_1$ is conserved, this is a contradiction.

%%%%%%%%%%%%%%%%%%%%%%%%%%%%%%%%%%%%%%%%%%%%%%%%%%%%%%%%%%%%
%%%%%%%%%%%%%%%%%%%%%%%%%%%%%%%%%%%%%%%%%%%%%%%%%%%%%%%%%%%%
\section{Dynamical epochs inside the horizon}

%%%%%%%%%%%%%%%%%%%%%%%%%%%%%%%%%%%%%%%%%%%%%%%%%%%%%%%%%%%%
\subsection{Simplified interior equations of motion}

Just like in recent AdS studies \cite{Hartnoll:2020rwq,Hartnoll:2020fhc,VandeMoortel:2021gsp}, 
we find that the dynamics inside the horizon of our asymptotically  flat hairy black holes separates into distinct epochs near the critical charge, $Q_c$ where the scalar field first turns on. These epochs are: the collapse of the Einstein-Rosen bridge, the Josephson oscillations and the Kasner epochs. We describe these in the next subsections.
Since we cannot solve the full equations exactly, we adopt the following strategy. We use the numerical solutions of the full equations to identify terms in the equations of motion which are negligible during the epoch of interest. We then drop those terms and find analytic solutions to the resulting equations. Finally, we compare the analytic solutions to the full numerical one.

During the ER collapse and Josephson epochs (and in some circumstances during the Kasner period), the strategy outlined in the previous paragraph indicates that we can neglect all the scalar field mass terms\footnote{Working with dimensionless quantities measured in scalar mass units, as we do, these are the two terms proportional to  $y_+^2=r_+^2 m^2$  in \eqref{eom}  that do not depend on any other parameters. 
}  in \eqref{eom} and the scalar charge term in \eqref{eomPhi}.
In these conditions, the equations of motion in the interior of the black hole are well approximated by 
\begin{subequations}\label{eominterior}
\begin{align}
& z^2 e^{-\frac{\chi}{2}}\left[\left(1+4 \alpha  \psi^2\right) e^{\frac{\chi}{2}} \Phi^\prime\right]^\prime\simeq 0 \quad \Leftrightarrow \quad \Phi^{\prime}\simeq\frac{E_0 e^{-\frac{\chi}{2}}}{1+4 \alpha  \psi^2}\,,\label{eominteriorPhi}
\\
%& z^2 e^{\frac{\chi}{2}} \left(\frac{e^{-\frac{\chi}{2}} f \psi^\prime}{z^2}\right)^\prime \simeq -\left(\frac{\tilde{q}^2 y_+^2 e^{\chi} \Phi^2}{f}+2 e^{\chi} z^2 \alpha  {\Phi^\prime}^2\right) \psi \,,\label{eominteriorPsi}
& z^2 e^{\frac{\chi}{2}} \left(\frac{e^{-\frac{\chi}{2}} f \psi^\prime}{z^2}\right)^\prime
\simeq -\left(\frac{\tilde{q}^2 y_+^2 e^{\chi} \Phi^2}{f} +      \frac{2 z^2 \alpha E_0^2 }{(1+4 \alpha  \psi^2)^2}\right) \psi \,,\label{eominteriorPsi}
\\
& \chi^\prime = 4\,z\,\left(\frac{\tilde{q}^2y_+^2 e^{\chi}}{f^2} \psi ^2 \Phi ^2+{\psi^\prime}^2\right)\,,
\label{eominteriorChi}
\\
&  \left(\frac{e^{-\frac{\chi}{2}} f}{z^3}\right)^\prime \simeq \left( \frac{E_0^2}{1+4 \alpha  \psi^2}- \frac{\kappa}{z^2}\right) e^{-\frac{\chi}{2}}  \,,\label{eominteriorf}
\end{align}
\end{subequations}
where $E_0$ is the constant electric field in the regimes where the approximations 
hold\footnote{As in \cite{Hartnoll:2020fhc}, the electric field is in the spacelike $t$ direction, while $\Phi$ is a component of the vector potential inside the horizon, with the $z$ coordinate being ‘time’. In the language of superconductors, the last term of the Maxwell equation \eqref{eomPhi} is a Josephson electric current in the interior, generated by  the condensate $\psi$ and vector potential  $\Phi$. We drop the Josephson current in \eqref{eominteriorPhi} since we observe (when comparing the approximate solution with the full numerical solution) that it does not backreact significantly on the electric field in any of the three epochs we will describe.}. 

\subsection{Collapse of the Einstein-Rosen bridge \label{sec:ER}}

As we increase the black hole charge slightly above the critical charge $Q_c$ for the scalar instability, the solution resembles RN until we approach the inner Cauchy horizon.
At this point, the scalar field triggers an instability very much like in recent AdS studies \cite{Hartnoll:2020rwq,Hartnoll:2020fhc}. This instability is stronger for small values of the scalar field, which highlights the nonlinear nature of the dynamics in this regime. The fundamental phenomenon, already observed in the AdS studies \cite{Hartnoll:2020rwq,Hartnoll:2020fhc}, is that as $g_{tt}$ approaches its would-be zero value at the Cauchy horizon, it suddenly suffers a very rapid collapse and becomes exponentially small. For this reason, in \cite{Hartnoll:2020rwq} this phenomenon was dubbed the collapse of the Einstein-Rosen (ER) bridge. This phenomenon has been recently established in the mathematical literature \cite{VandeMoortel:2021gsp}. 

As in \cite{Hartnoll:2020rwq,Hartnoll:2020fhc}, for very small scalar field the instability is so fast that we can keep the $z$ coordinate  essentially fixed. This means that we can take $z = z_\star + \delta z$, so that all explicit factors of $z$ in the equations of motion are set to the constant $z_\star$ and $f,\chi,\psi,\Phi$ are now functions of $\delta z$.  $z_\star$ is close to the inner horizon of RN. Moreover, as the numerical solution of the full equations confirms, in this limit the Maxwell potential $\Phi$ is large compared to its derivative $\Phi^{\prime}\propto E_0$. It follows that in the right hand side of \eqref{eominteriorPsi} we can neglect the second term ($\alpha \Phi^{\prime \,2}\,\propto\,  \alpha E_0^2$) when compared to the first term ($\propto \Phi^2$)  and in \eqref{eominteriorPsi} and \eqref{eominteriorChi} we can set $\Phi\simeq \Phi_o$. However, comparison with the numerical full solution indicates that it is fundamental to keep the term proportional to $\Phi^\prime \propto E_0$ in \eqref{eominteriorf}, although we can do the approximation $1+4\alpha \psi^2\simeq 1$ in this term since the scalar field $\psi$ is very small. Altogether,  the equations of motion read,
\begin{subequations}\label{EReom}
\begin{align}
&  \Phi\simeq \Phi_o\,,\label{EReomPhi}
\\
&  \left(e^{-\frac{\chi}{2}} f \psi^\prime\right)^\prime
 + \frac{\tilde{q}^2 y_+^2 e^{\frac{\chi}{2}} \Phi_o^2}{f} \psi \simeq 0\,,\label{EReomPsi}
\\
& \chi^\prime \simeq 4\,z_\star\,\left(\frac{\tilde{q}^2y_+^2 e^{\chi}\Phi_o^2}{f^2} \psi ^2 +{\psi^\prime}^2\right)\,,
\label{EReomChi}
\\
&  \left(e^{-\frac{\chi}{2}} f \right)^\prime \simeq z_\star \left( E_0^2z_\star^2 - \kappa \right) 
e^{-\frac{\chi}{2}} \,.\label{EReomf}
\end{align}
\end{subequations}
With these approximations (validated \`a posteriori) the solution in the collapse of the ER bridge is very similar to the one found in the AdS system of \cite{Hartnoll:2020fhc} with the cosmological constant term replaced by the $S^2$ curvature contribution proportional to  $\kappa=1$ in \eqref{EReomf}. In particular, note that  Maxwell-scalar coupling terms proportional to $\alpha$ do not appear in \eqref{EReom}.        

\begin{figure}[h!]
    \centering
    \vspace{1cm}
    \includegraphics[width=\textwidth]{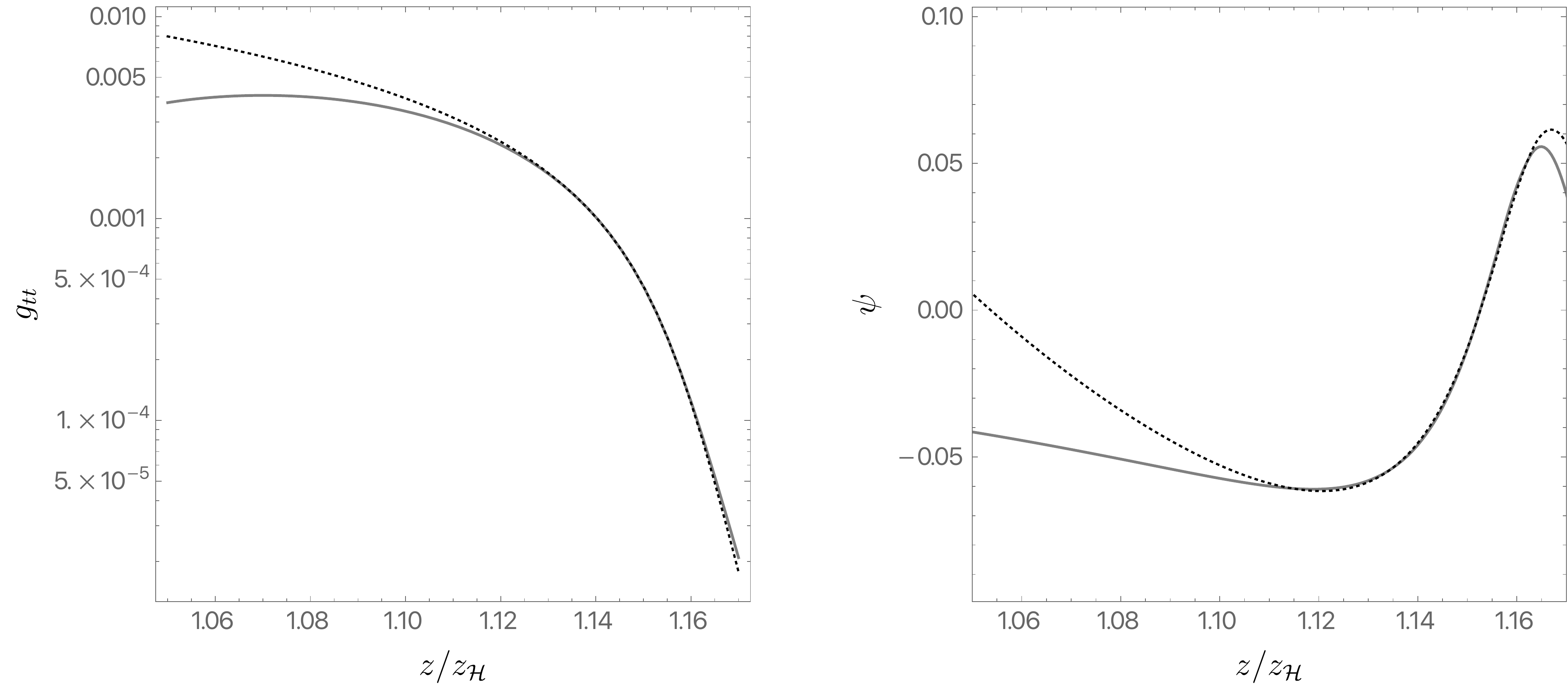}
    \caption{A comparison of the numerical solutions (solid grey lines) and fits to the analytic expressions \eqref{ER:psi} and \eqref{eq:kink} (black dotted curves) describing the collapse of the Einstein-Rosen bridge. This is for $m M=0.941632$, $Q/Q_c = 1.001833$, $q=m$ and $\alpha=1$.}
       \label{fig.ercollapse}
\end{figure}

Equation \eqref{EReomPsi} can be solved explicitly yielding
\be\label{ER:psi}
\psi \simeq \psi_o \cos \left(y_+\tilde{q} \Phi_o \int_{z_\star}^z \frac{e^{\chi/2}dz}{f} + \varphi_o \right) \,,
\ee
where $\psi_o$ and $\varphi_o$ are two integration constants.
%Note that if $f$ develops a zero, then $f\sim f'(z-z_{\mathcal{I}})$ and one has the expected logarithmic oscillations near to the inner horizon $z=z_{\mathcal{I}}$ observed in studies of the linear instability of this horizon.
To get the gravitational field, one first observes that, interestingly, the scalar field oscillations \eqref{ER:psi} drop out of the equation \eqref{EReomChi} for $\chi$. This is because inserting \eqref{ER:psi} on the right hand side of \eqref{EReomChi} one finds that it reduces to a constant,  $f^{-2}\tilde{q}^2y_+^2 e^{\chi}\Phi_o^2\psi_o^2$. 
Now, it is useful to recall that, from \eqref{ansatz}, one has $g_{tt}=\frac{1}{z_\star^2} fe^{-\chi}$. Taking the derivative of  \eqref{EReomf}, and after some algebra that uses the definition of $g_{tt}$, its derivative and \eqref{EReomf} itself, we find that $g_{tt}$ must obey the second order nonlinear ODE
\be\label{Joseph:gtt}
\frac{g_{tt}^{\prime\prime}}{g_{tt}^{\prime}}-\frac{c_1^2 g_{tt}^{\prime}}{g_{tt}(c_1^2+g_{tt})}\simeq 0 \,, \qquad \hbox{with}\quad c_1^2 = \frac{2 y_+^2 \tilde{q}^2  \Phi_o^2\psi_o^2}{z_\star^2( E_o^2z_\star^2-\kappa)} \,.
\ee
Apart from the particular value of the constant $c_1$, this is the same ODE found in the collapse of the ER bridge of the AdS studies \cite{Hartnoll:2020rwq,Hartnoll:2020fhc} and thus it has a similar solution:
\be\label{eq:kink}
c_1^2 \ln(g_{tt}) + g_{tt} = - c_2^2 (z - z_o) \quad \Leftrightarrow \quad 
g_{tt}=c_1^2 \,W\left( c_1^{-2} e^{-(c_2/c_1)^2\, \left(z-z_o \right)}\right)
\,,
\ee
where $c_2 > 0$ and $z_o$ are integration constants and $W(x)\equiv ProducLog(x)$  gives the principal solution for $w$ in $x=w e^w$. For $z < z_o$, $g_{tt} \propto (z_o - z)$ is linearly vanishing, as in the approach to an inner horizon, but for $z > z_o$, $g_{tt} \propto e^{-(c_2/c_1)^2 (z - z_o)}$ is nonzero but exponentially small, instead of vanishing or changing sign. This collapse occurs over a coordinate range $\Delta z = (c_1/c_2)^2$. Since $c_1$ is proportional to $\psi_0$ which vanishes as $Q\rightarrow Q_c$, this justifies our assumption that the collapse happens very quickly after the scalar field turns on. In particular, the radius of the transverse spheres barely changes.
These solutions agree well with the full numerical evolution as shown in Fig.~\ref{fig.ercollapse}. Having $g_{tt}$ we can now get the solution for $\chi$ and $f$: 
\be\label{ER:fChi}
e^{-\chi}=\frac{c_2^2 g_{tt}^2}{\left(c_1^2+g_{tt}\right)^2} \frac{z_\star}{E_0^2 z_\star^2-\kappa}\,, \qquad f=-z_\star^2 e^{\chi}  g_{tt}\,,
\ee
 where we used the fact that $g_{tt}^\prime=-\frac{c_2^2 g_{tt}}{c_1^2+g_{tt}}$ which follows from \eqref{eq:kink}.

%%%%%%%%%%%%%%%%%%%%%%%%%%%%%%%%%%%%%%%%%%%%%%%%%%%%%%%%%%%%
 \subsection{Josephson oscillations \label{sec:Josephson}}

Although we have neglected the Josephson current term (i.e. the last term in \eqref{eomPhi}) in the interior equation \eqref{eominteriorPhi}, the scalar field solution \eqref{ER:psi} encodes information about Josephson oscillations. Indeed, inside the horizon $z$ is a timelike coordinate while $t$ is spacelike. Moreover, the argument of the cosine in \eqref{ER:psi} can be written as $y_+\tilde{q} \int A_{\hat t} \, d\tau$, where  $d\tau = \sqrt{g_{zz}} dz$ is  the proper time and $A_{\hat t} = A_t/\sqrt{g_{tt}}$ is the vector potential in locally flat coordinates. A nonzero $A_{\hat t}$ indicates a phase winding in the $t$ direction. The scalar condensate $\psi$ determines the superfluid stiffness. Thus,  \eqref{ER:psi} describes oscillations in time of the superfluid stiffness sourced by a background phase winding, a phenomenon that is known as the Josephson effect. 
In the aftermath of the collapse of the ER bridge, these Josephson oscillations become (for $Q \approx Q_\text{c}$) the dominant feature in the solution over a regime that is naturally denoted as the Josephson oscillations epoch and that we now describe.  

By the end of the collapse of the ER bridge epoch, the derivative of the Maxwell field $\Phi' $ is still very small. This is essentially because $\Phi' \propto e^{-\chi/2}$ $-$ see \eqref{eominteriorPhi} $-$ and $e^{-\chi/2}$ is very small. It follows that in the Josephson oscillation epoch we can still 
neglect the $E_0 e^{-\frac{\chi}{2}}$  contribution in \eqref{eominteriorPhi} but we can also take $\left( \frac{E_0^2}{1+4 \alpha  \psi^2}- \frac{\kappa}{z^2}\right) e^{-\frac{\chi}{2}} \simeq 0$  in the interior equation \eqref{eominteriorf}. Thus, the Maxwell field $\Phi$ and gravitational field $f$ are given by 
\be\label{eq:pos}
 \Phi \simeq \Phi_o  \,, \qquad \frac{f e^{-\chi/2}}{z^3} \simeq - \frac{1}{c_3} \,,
\ee
with $c_3$ constant. To determine the latter, we match the Josephson oscillation solution \eqref{eq:pos} with the $z > z_o$ solution \eqref{ER:fChi} of the collapse of the ER bridge in the region where they overlap. This yields
  \be\label{c3}
c_3\simeq \frac{c_2}{c_1^2}\sqrt{\frac{z_\star^3}{E_0^2 z_\star^2-\kappa}} \,,
\ee
where we have approximated $c_1^2+g_{tt}^2(z_\star) \simeq c_1^2$ since  $g_{tt}$ is very small near the inner horizon.  It follows that $c_3$ becomes large as $Q \to Q_\text{c}$. 

Inserting \eqref{eq:pos} into \eqref{eominteriorPsi} the scalar field must obey the Bessel equation $\left(z\psi^\prime\right)^\prime-c_3^2 y_+^2\tilde{q}^2\Phi_o^2 z^{-5}\psi\simeq 0$ whose solution is
\be\label{Joseph:psi}
\psi \simeq c_4 J_0\left(\frac{y_+|\tilde{q} \Phi_o| c_3}{2 z^2}\right) + c_5  Y_0\left(\frac{y_+ |\tilde{q}  \Phi_o| c_3}{2 z^2} \right) \,,
\ee
where $c_4$ and $c_5$ are integration constants. 
To find these constants we need to match the small $z$ behaviour of the Josephson solution \eqref{Joseph:psi} and its derivative $\psi^\prime$ with the large $z$ behaviour of the  ER bridge collapse solution \eqref{ER:psi} and its derivative in the overlapping region around $z_\star$. This yields 
\begin{subequations}\label{c4c5} 
\begin{align}
& c_4 \simeq \left(\frac{z_\star \pi^2 c_2^2}{32} \, \frac{\psi_o^2}{c_1^2} \right)^{1/4} \sin \left(\frac{c_2 \sqrt{z_\star}}{\sqrt{2}}\, \frac{1}{\psi_o c_1} - \varphi_o + \frac{\pi}{4} \right)\,,
\\
& c_5 \simeq \left(\frac{z_\star \pi^2 c_2^2}{32} \, \frac{\psi_o^2}{c_1^2} \right)^{1/4} \sin \left(\frac{c_2 \sqrt{z_\star}}{\sqrt{2}}\, \frac{1}{\psi_o c_1} - \varphi_o - \frac{\pi}{4} \right)\,, 
\end{align}
\end{subequations}
So the oscillations of the scalar field start in the collapse of the ER bridge regime $-$ see \eqref{ER:psi} $-$ and they propagate continuously onto 
the Josephson oscillation epoch where they are described by the Bessel oscillation \eqref{Joseph:psi}. As $c_3$ is large these oscillations are very fast. These oscillations will  propagate further into the interior so it is important to study the large $z$ behavior of the scalar field \eqref{Joseph:psi}:
\be\label{eq:tokasner}
\psi|_{\hbox{\tiny large} \, z} \simeq \frac{2 c_5}{\pi} \ln \left(c_3 \frac{y_+\tilde{q} e^{\gamma_E} \Phi_o}{4 z^2} \right) + c_4 + \cdots \,,
\ee
with $\gamma_E$ being the Euler-Mascheroni constant. The logarithmic behavior indicates the onset of a Kasner regime, that we will describe in the following section. 
%The constant $c_5$ in \eqref{eq:tokasner} will determine the Kasner exponent $\beta$ (to be introduced later). 

\begin{figure}[ht!]
    \centering
    \vspace{1cm}
    \includegraphics[width=\textwidth]{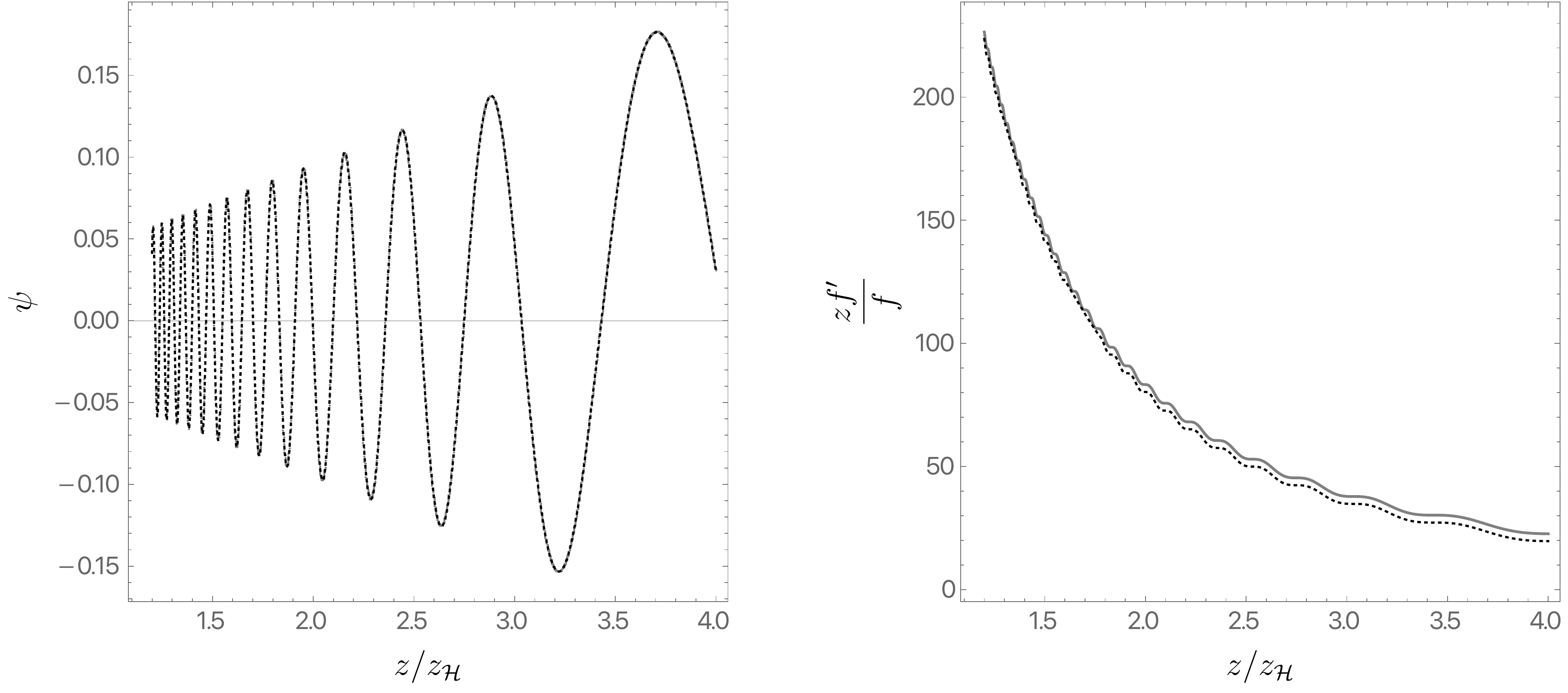}
    \caption{A comparison of the numerical solutions (solid grey lines) and fits to the analytic expressions \eqref{Joseph:psi} and \eqref{Joseph:chif} (black dotted curves) describing Josephson oscillations. This is for $m M=0.941632$, $Q/Q_c = 1.001833$, $q=m$ and $\alpha=1$.}
       \label{fig.josephson}
\end{figure}

To complete our discussion of the scalar field during the Josephson epoch, in the left panel of Fig.~\ref{fig.josephson} we compare the analytical approximation \eqref{Joseph:psi} (black dotted line) with the numerical data for the scalar field (continuous line) for $\alpha = 1, q = m, mM = 0.941632$ and $Q/Q_c=1.001833$. Again, we confirm that the assumptions made to find the analytical description of the system during the intermediate Josephson epoch is in excellent agreement (as $Q\to Q_c$ where $\psi$ is small) to the  numerical solution of the full equations of motion \eqref{eom}.  

Finally, we can also find the metric solution. One first notes that inserting \eqref{eq:pos} into \eqref{eominteriorChi} yields the equation $\chi^\prime = 4\left(z{\psi^\prime}^2+ z^{-5}c_3^2 y_+^2 \tilde{q}^2 \Phi_o^2 \psi^2 \right)$ that we can solve for $\chi$. We can insert this back into \eqref{eq:pos} to get the solution for $f$. Altogether we find that 
\begin{subequations}\label{Joseph:chif}
\begin{align}
& \chi\simeq 2\ln(f_o c_3)+4 \int_{z_\star}^z \left[\tilde{z} \psi^{\prime\,2} + \frac{y_+^2\tilde{q}^2 \Phi_o^2 c_3^2 \psi^2}{\tilde{z}^5} \right] d\tilde{z}\,,
\\
& f \simeq - f_o z^3 \exp \left\{2\int_{z_\star}^z \left[\tilde{z} \psi^{\prime\,2} + \frac{y_+^2\tilde{q}^2 \Phi_o^2 c_3^2 \psi^2}{\tilde{z}^5} \right] d\tilde{z} \right\} \,, 
\end{align}
\end{subequations}
where $f_o$ is a constant of integration, and $g_{tt}=z^{-2} fe^{-\chi}$. The Bessel solution \eqref{Joseph:psi} should be inserted into these integrals. The integrals can be done analytically in terms of Bessel functions.
These describe the small oscillations seen in $z f'/ f$ in the right panel of Fig.~\ref{fig.josephson}. Altogether, the two plots of Fig.~\ref{fig.josephson} verify that, at small $Q-Q_c$, the functional forms \eqref{Joseph:psi} and (\ref{Joseph:chif})  fit the numerical solutions to the full differential equations \eqref{eom} all the way from the end of the ER collapse, through the Josephson oscillation epoch, and  till the beginning of the subsequent Kasner regime.

Note that, like the collapse of the ER bridge epoch, the solution in the Josephson oscillations region is  similar to the one found in the AdS system of \cite{Hartnoll:2020fhc} (with the cosmological constant term now replaced by the $S^2$ curvature contribution). In particular, the Maxwell-scalar coupling terms proportional to $\alpha$ are again not relevant. This will no longer be the case for the Kasner transitions that we discuss next.

%%%%%%%%%%%%%%%%%%%%%%%%%%%%%%%%%%%%%%%%%%%%%%%%%%%%%%%%%%%%
\subsection{Kasner epochs and transitions \label{sec:Kasner}} 

As shown  in \eqref{eq:tokasner}, at the end of the Josephson oscillations, $\psi$ grows logarithmically. This marks the entrance to a new era described by a Kasner cosmology as we now explain.
If we write $\psi = \beta \ln z$ with 
\be\label{def:beta}
\beta = -\frac{4}{\pi}\,c_5\,, 
\ee
and plug it into \eqref{Joseph:chif} we find that the large $z$ behavior of $\chi$ and $f$ is
\be 
f \simeq - f_o z^{3+2\beta^2}, \qquad
\chi \simeq 4\beta^2 \ln z\,.
\ee
This corresponds to a metric in which all components are powers of $z$. Furthermore, from \eqref{eominteriorPhi}, the Maxwell potential is
\be \label{MaxKasner}
\Phi \simeq \Phi_K + \frac{E_K\ z^{1-2\beta^2}}{1+4\alpha\beta^2\ln^2 z}\,.
\ee
So, as long as $\beta^2 > 1/2$, the Maxwell field will remain unimportant at large $z$. (We will consider the consequences of $\beta^2 < 1/2$ later.)
Introducing a proper time $\tau=\int \sqrt{g_{zz}}\mathrm{d}z\ \propto \ z^{-(3/2 + \beta^2)}$ (so $z=\infty$ corresponds to $\tau=0$) and setting $\kappa = 0$ (since the curvature of the sphere is negligible in this regime), the solution takes the standard Kasner form \cite{KasnerGeometricalTO,Belinski:1973zz}
\be \label{KasnerCosmo}
ds^2 = -d\tau^2 + \tau^{2p_t} dt^2 + \tau^{2p_x} (dx^2 + d\phi^2),\qquad\psi = p_\psi \ln \tau
\ee
with 
\be \label{KasnerCosmoExp}
p_t = \frac{2\beta^2 -1}{2\beta^2 +3}, \qquad p_x = \frac{2}{2\beta^2 + 3}, \qquad p_\psi = \frac{2\beta}{2\beta^2 + 3}\,.
\ee
These exponents satisfy the usual Kasner relations\footnote{The factor of $4$ in front of $p_\psi^2$ comes from our normalization of $\psi$ in the action.}
\be \label{KasnerCosmoExp2}
p_t + 2p_x = 1, \qquad p_t^2 + 2p_x^2 + 4 p_\psi^2 = 1\,.
\ee
The metric \eqref{KasnerCosmo} has a spacelike curvature singularity at $\tau = 0$ ($z = \infty$) in all cases except $p_t = 1$, which corresponds to $\beta = \infty$. 

\begin{figure}[h!]
    \centering
    \vspace{1cm}
    \includegraphics[width=0.9\textwidth]{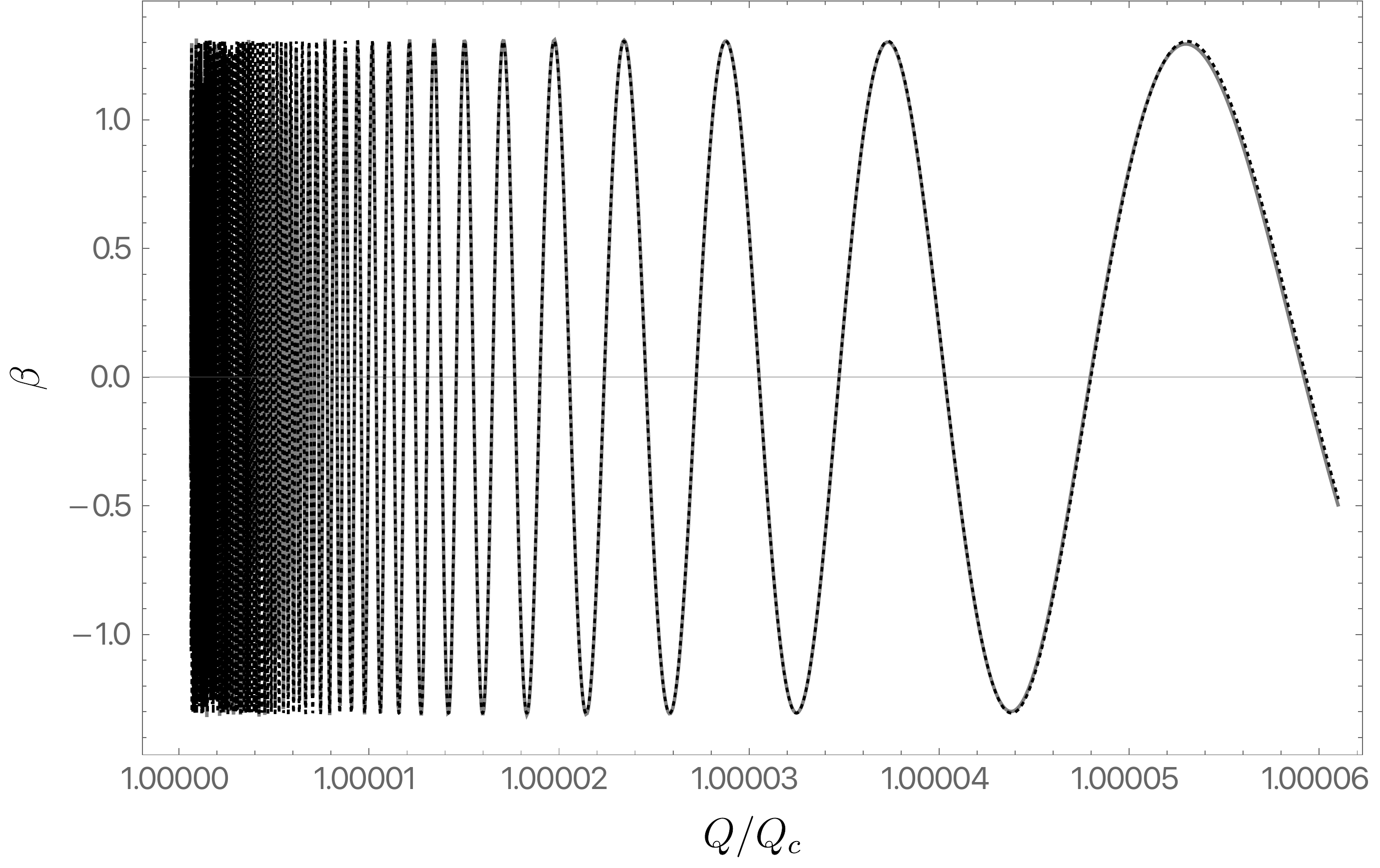}
    \caption{The parameter $\beta$ determining the Kasner exponents as a function of the black hole charge $Q$ for  $\alpha = 1, q = m, mM = 0.5$. The solid grey line is the numerical solution and the dotted black curve is a fit to \eqref{eq:osc}. The scalar starts to condense at the critical charge $m\,Q_c = .4942545$.}
       \label{fig.osc}
\end{figure}

From \eqref{def:beta}, the parameter $\beta$ controlling the Kasner exponents is proportional to $c_5$, which from \eqref{c4c5} is an oscillating function of the parameters. Numerically, we find that near $Q_c$, $\beta$ is very well fit by
\be \label{eq:osc} 
\beta = A\sin \left[ \frac{B}{Q/Q_c -1} + C \right]  
\ee
over many oscillations. This is demonstrated in Fig.~\ref{fig.osc} where we see that the black dotted line describing \eqref{eq:osc}  with $A\simeq 1.32104(2)$, $B\simeq 7.90056(3)\times 10^{-4}$ and $C\simeq -0.73961(2)$ (for  $\alpha = 1, q = m, mM = 0.5$) is in excellent agreement with the numerical data (black continuous line) over many oscillations for small $Q-Q_c$. This clearly shows the extreme sensitivity of the Kasner exponents on the charge near the critical charge $Q_c$.

We can also find a good analytical approximation for the amplitude of the $\beta$ oscillations near the critical  charge $Q_c$. Indeed, as $Q\to Q_c$ the following approximations should be excellent: 1) $z_0\sim z_\star \sim z_{\mathcal{I}} =1/\mu^2$, 2) $c_1$, being proportional to the scalar condensate
 as described by \eqref{Joseph:gtt}, approaches  $c_1\propto \psi_0 \to 0$ which allows us to find $c_2\simeq y_+\sqrt{1-\mu^2}$ via \eqref{eq:kink} evaluated at $z=z_{\mathcal{I}}$ (i.e. $c_2$ attains its RN value), and 3) $\Phi_0=\Phi(z_{\mathcal{I}})\simeq -1/\mu$ and $E_0=\Phi^\prime(z_{\mathcal{I}})\simeq -\mu$ are effectively also given by their RN values as read from \eqref{RNsol:int}. 
 We can now insert these quantities in \eqref{c4c5} and \eqref{def:beta} to obtain that as $Q\to Q_c$, the amplitude in \eqref{eq:osc} should be well approximated by
 \be\label{betaOsc:Amp}
A^4 \simeq \frac{4}{\pi ^2 y_+^2 \tilde{q}^2 \mu^6}\,.
\ee 
For  $\alpha = 1, q = m, mM = 0.5$, this gives $A\simeq 1.32166(3)$ which is in very good agreement with the numerical value of $A\simeq 1.32104(2)$.

So far we have described the system at the end of the Josephson regime when it moves to the Kasner epoch. Next, we would like to understand what happens when the system evolves further inside the Kasner epoch. From \eqref{MaxKasner}, one sees that when $\beta^2 > 1/2$ at the beginning of the Kasner epoch, the Maxwell field remains bounded for large $z$. We then expect (and confirm it to be so) that in such cases, as we enter deeper into the Kasner regime all the way to the singularity at $z\to \infty$, the system remains described by the Kasner cosmology \eqref{KasnerCosmo}-\eqref{KasnerCosmoExp} with the same exponents. 

On the other hand, when $\beta^2 < 1/2$ at the beginning  of the Kasner epoch, there are new effects. Indeed, the growing
 Maxwell field \eqref{MaxKasner} causes a transition  to a different Kasner solution with new exponents.\footnote{Similar transitions were found inside Einstein-Yang-Mills black holes \cite{Donets:1996ja,Breitenlohner:1997hm}.} This transition  can be described as follows.
Numerically, it is found that once the solution enters the Kasner epoch, the curvature, mass and charge terms all become  negligible. Thus the interior equations of motion  \eqref{eominterior} further simplify to
\begin{subequations}
\label{eomapp}
\begin{align}
&\left[\left(1+4 \alpha  \psi^2\right) e^{\frac{\chi}{2}} \Phi^\prime\right]^\prime \simeq 0 
\quad \Leftrightarrow \quad \Phi^{\prime}\simeq\frac{E_0 e^{-\frac{\chi}{2}}}{1+4 \alpha  \psi^2} \,, \label{eomappPhi}
\\
& z^2 e^{\frac{\chi}{2}} \left(\frac{e^{-\frac{\chi}{2}} f \psi^\prime}{z^2}\right)^\prime+2e^{\chi} z^2 \alpha  {\Phi^\prime}^2 \psi\simeq 0\,,\label{eomappPsi}
\\
& \chi^\prime-4\,z\,{\psi^\prime}^2\simeq 0\,,\label{eomappChi}
\\
& e^{\frac{\chi}{2}} z^4 \left(\frac{e^{-\frac{\chi}{2}} f}{z^3}\right)^\prime-\left(1+4 \alpha  \psi^2\right) e^{\chi} z^4 {\Phi^\prime}^2\simeq 0\,.\label{eomappf}
\end{align}
\end{subequations}
Quite importantly, unlike in the previous two epochs, this time we cannot neglect the terms proportional to the Maxwell-scalar coupling $\alpha$ in \eqref{eomapp}. Moreover, unlike the Josephson epoch where the Maxwell field was approximately constant, now the Maxwell field variation will have an important contribution to the evolution of the system. 

The first and last equations of \eqref{eomapp} can be readily integrated 
if we note that they are the homogeneous Josephson equations, with solution \eqref{eq:pos}, but this time with a small source added. It follows that the solution of the non-homogeneous \eqref{eomappPhi} and \eqref{eomappf} is given by the homogeneous solution \eqref{eq:pos} plus the particular integral of the source:
\begin{subequations}
\begin{align}
&\Phi \simeq \Phi_o+E_0 \int \frac{e^{-\frac{\chi}{2}}}{1+4 \alpha \psi^2}\mathrm{d}z\,,
\\
&f \simeq e^{\frac{\chi}{2}}z^3 E_0^2\left(\int \frac{e^{-\frac{\chi}{2}}}{1+4 \alpha \psi^2}\mathrm{d}z-\frac{1}{c_3}\right), \label{eomappf:soln}
\end{align}
\end{subequations}
where $c_3$ is given by \eqref{c3}.

The remaining two equations \eqref{eomappPsi}-\eqref{eomappChi} can now be combined to form a third order differential equation for $\psi$. For that, we take the derivative of \eqref{eomappPsi} and, after some algebra that uses \eqref{eomappPhi}, \eqref{eomappChi}, \eqref{eomappf} and \eqref{eomappf:soln}  to eliminate $\Phi',\chi', f'$ and $f$, we find that $\psi(z)$ obeys the third order ODE: 
\begin{multline}\label{eq:3rdODE}
1+\left(1+4 \alpha  \psi^2\right) \left[\frac{1}{1+4 \alpha  \psi^2}\frac{1}{\psi^\prime+z \psi^{\prime\prime}}\left(\psi^\prime z+\frac{2 \alpha \,\psi}{1+4 \alpha \,\psi^2}\right)\right]^\prime
\\- \frac{2 z {\psi^\prime}^2}{\psi^\prime+z \psi^{\prime\prime}}\left(\psi^\prime z+\frac{2 \alpha \,\psi}{1+4 \alpha \,\psi^2}\right)\simeq 0\,.
\end{multline}
We can convert this to a second order equation using the following change in variables
\begin{equation}\label{def:S}
\psi^\prime(z) = \frac{S(\psi(z))}{z}.
\end{equation}
 After some lengthy algebra we find
%\begin{multline}
%\frac{\mathrm{d}}{\mathrm{d} \psi}\left(Q \frac{\mathrm{d}Q}{\mathrm{d} \psi}\right)-\Bigg\{1-2 Q^2+2 \left[1-\frac{2 \alpha \psi }{Q+2 \alpha  \psi  (1+2 \psi  Q)}\right] \frac{\mathrm{d}Q}{\mathrm{d} \psi}+
%\\
%\frac{2 \alpha Q}{Q+2 \alpha  \psi  (1+2 \psi  Q)}\left(1-\frac{16 \alpha  \psi ^2}{1+4 \alpha  \psi ^2}-4 \psi  Q\right) \Bigg\}\frac{\mathrm{d}Q}{\mathrm{d} \psi}=0\,.
%\label{eq:S}
%\end{multline}
\begin{multline}\label{eq:S}
S\frac{\mathrm{d}}{\mathrm{d}\psi}\left\{\frac{1}{1+4\alpha \psi^2}\left[1+\frac{2\alpha \psi}{(1+4 \alpha \psi^2)S}\right]\frac{1}{\dot{S}}\right\}
\\+\frac{1}{1+4\alpha \psi^2}+\frac{1-2 S^2}{\dot{S}(1+4 \alpha \psi^2)}\left[1+\frac{2\alpha \psi}{(1+4 \alpha \psi^2)S}\right]\simeq 0\,,
\end{multline}
where $\dot{}$ means derivative with respect to $\psi$. Finding $\psi(z)$ now amounts to first solving this second order ODE for $S(\psi)$ and then doing the simple integration \eqref{def:S}. 

As a warm-up exercise to understand the solutions of \eqref{eq:S}, consider first the limit  $\alpha=0$. In this case we find that \eqref{eq:S} admits an analytic solution of the form
\begin{equation}\label{Qsoln0}
S^{\alpha=0}(\psi)=\frac{1+2 \beta _o^2}{4 \beta _o}+\frac{\left(1-2 \beta _o^2\right)}{4 \beta _o} \tanh \left[\frac{\left(1-2 \beta _o^2\right) \left(\psi -\psi_o\right)}{2 \beta _o}\right]\,.
\end{equation}
This solution has the same functional form as its AdS counterpart of \cite{Hartnoll:2020fhc}. In particular, for $\psi\to-\infty$, $S^{\alpha=0}$ is a constant. Indeed, assuming $|\beta_o|\leq 1/\sqrt{2}$, we have set our integration constants in a such a way that
\begin{equation}\label{psiBeforeInv0}
\lim_{\psi \to-\infty} S^{\alpha=0}(\psi)= \beta_o\,,
\end{equation}
which defines $\beta_o$.
 Taking the opposite limit of large $\psi$, yields
\begin{equation}\label{psiAfterInv0}
\lim_{\psi \to+\infty} S^{\alpha=0}(\psi)= \frac{1}{2\beta_o}\,,
\end{equation}
which is exactly the Kasner inversion  found previously in AdS  \cite{Hartnoll:2020fhc}. That is to say, if the Kasner transition occurs near $z_\mathrm{tr}$, and  one has $|\beta|< 1/\sqrt{2}$ and thus growing $g_{tt}$ for $z\ll z_\mathrm{tr}$,
the Maxwell field growth \eqref{MaxKasner}  triggers a transition to a new Kasner regime with exponent $\beta_\mathrm{new}=1/(2\beta)$ with $|\beta_\mathrm{new}|> 1/\sqrt{2}$ for $z\gg z_\mathrm{tr}$, and thus with decreasing $g_{tt}$.

The $\alpha$-dependent terms will modify this behavior. However, it seems difficult, if not impossible,  to solve \eqref{eq:S} analytically for generic values of $\alpha$. If $\alpha$ is small, we can do perturbation theory by setting
\begin{equation}
S=S^{\alpha=0}+\alpha \tilde{S}\,,
\end{equation}
and expanding around $\alpha=0$. This exercise turns out to be doable, though the expression for $\tilde{S}$ involves $\mathrm{PolyLog}$ functions and is rather cumbersome. Yet, we can obtain the limits $\psi\to \pm \infty$ which allow us to extract the relevant information for our purposes. We find that if
\begin{equation}\label{psiBeforeInv}
\lim_{\psi \to-\infty} S(\psi)= \beta_o\,,
\end{equation}
then
\begin{equation}\label{psiAfterInv}
\lim_{\psi \to+\infty} S(\psi)= \frac{1}{2\beta_o}-\left[\psi_o+\frac{\beta _o \log \left(2 \beta _o^2\right)}{1-2 \beta _o^2}\right]\frac{1+2 \beta _o^2}{\beta _o^2}\alpha\,.
\end{equation}
This results shows two things: 1) unlike the $\alpha=0$ case, the behaviour at large $\psi$ now depends not only on $\beta_o$ but also on $\psi_o$, and 2) perturbation theory will break down when $\beta_o \rightarrow 0$.

\begin{figure}[h!]
    \centering
    %\vspace{1cm}
    \includegraphics[width=0.9\textwidth]{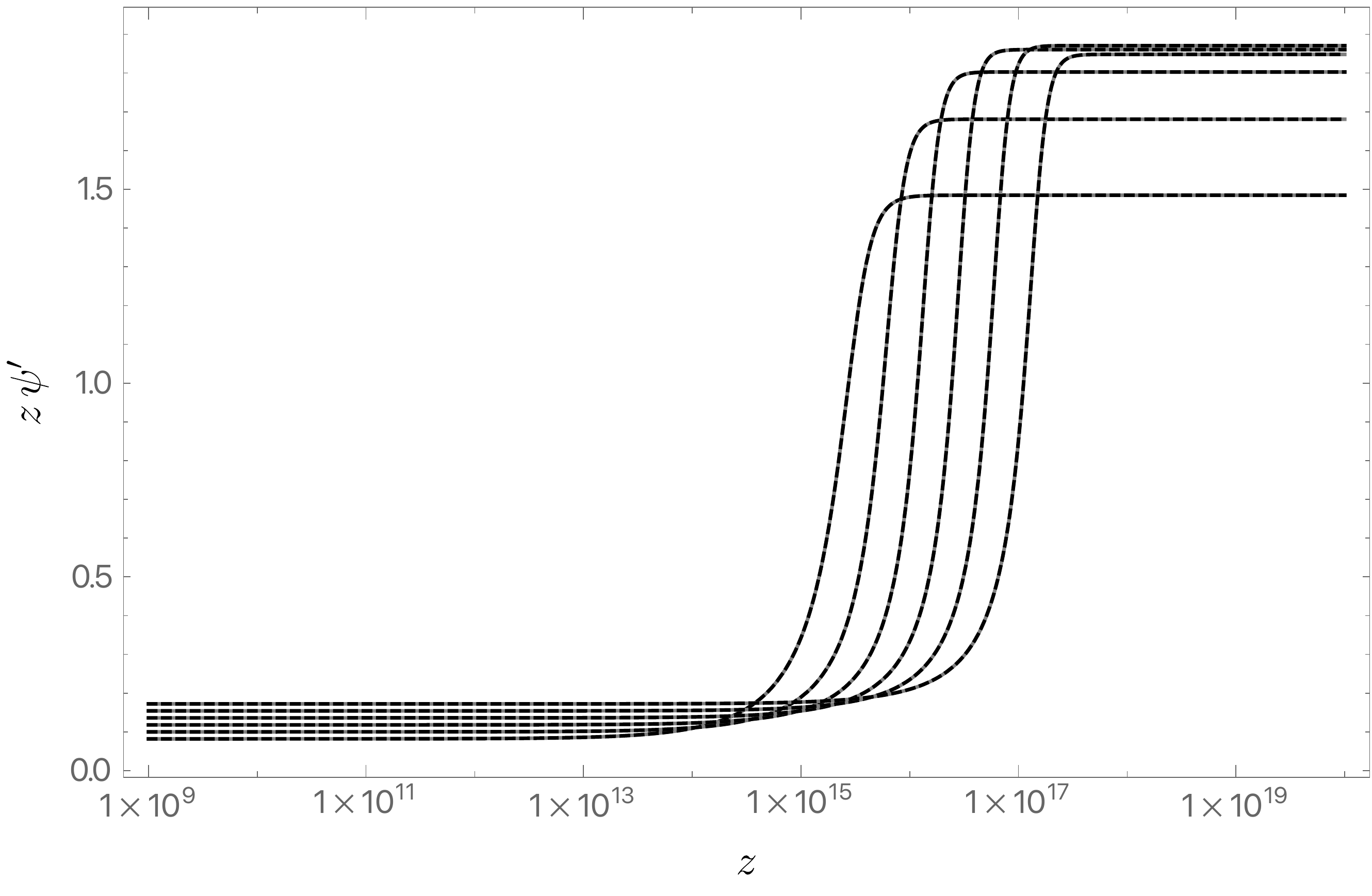}
    \caption{Kasner transitions between $S|_{z\ll z_\mathrm{tr}}=\beta$ and $S|_{z\gg z_\mathrm{tr}}=\beta_\mathrm{new}$ where $z_\mathrm{tr}$ is the location of the transition. This is for $\alpha = 1, q =m, mM =0.5$ and the different curves have different $\beta$, namely (from bottom to top on the left side): $\beta=\{0.82056\,,1.0004\,,1.1804\,,1.3606\,,1.5409\,,1.7212\}\times 10^{-1}$ and $Q/Q_c-1=\{5.8918,5.8857,5.8796,5.8735,5.8674,5.8613\}\times 10^{-5}$. For each curve, the gray solid line is the numerical solution of \eqref{eom} and the dashed black line is the solution of the approximated ODE \eqref{eq:3rdODE}: they are are superposed.}
       \label{fig.KasnerInversions}
\end{figure}

Having exhausted our attempts to solve analytically, for generic and finite $\alpha$, the approximate ODE for the scalar field \eqref{eq:S} or \eqref{eq:3rdODE} describing a Kasner transition, we must resort to numerical solutions of \eqref{eq:3rdODE}. Even though we have numerical solutions of the full set of equations \eqref{eom}, it is still of interest to verify whether the single equation \eqref{eq:3rdODE} captures all the important effects.
For that, we first choose a value $z=z_c$ in the middle of the first Kasner epoch and compute
%identify the value $z=z_c$ at which the first Kasner regime starts (this coincides with the minimum value of $\psi$ after the Josephson oscillations, i.e. at the beginning of the Kasner epoch)
 $\psi,\psi^\prime,\psi^{\prime\prime}$ at $z=z_c$. Equation \eqref{eq:3rdODE} can then be integrated numerically  to larger values of $z$ to obtain $S(z)=z \psi^\prime(z)$.
Once this integration is completed, we can check whether the profile $S(z)$ obtained from solving \eqref{eq:3rdODE}  matches the numerical solution of the full equation of motion \eqref{eom}. Fig.~\ref{fig.KasnerInversions} which plots $S=z\psi^\prime$ as a function of $z$ demonstrates that this is indeed the case. 
When $S$ is constant in each of the curves, it describes a Kasner epoch with $S = \beta $ on the left side and $S=\beta_\mathrm{new}$ on the right side. Different curves correspond to different starting values of $0<\beta<1/\sqrt{2}$, as identified in figure's caption, and after the transition all curves have $\beta_\mathrm{new}>1/\sqrt{2}$. 
In each of these curves, we see that \eqref{eq:3rdODE} indeed  describes the Kasner transitions very well, since the numerical integration of \eqref{eq:3rdODE} (dashed black line) is essentially on top of the gray solid line that solves the full coupled ODE system \eqref{eom}. For this figure, we have set $\alpha = 1$, and the $\alpha$ dependent terms are important to describe this transition.
In Fig.~\ref{fig.KasnerInversions} we also observe an interesting property. For the two upper curves, the curve starting with higher $\beta$ has lower $\beta_\mathrm{new}$, very much like the Kasner inversion observed in the AdS case of \cite{Hartnoll:2020fhc}. However, in the four lowest curves we find that
 increasing the initial $\beta$ also increases the final $\beta_\mathrm{new}$. This is why we are calling this a Kasner transition, rather than an inversion.

To recap, after the Josephson oscillations the solution enters a Kasner epoch described by \eqref{KasnerCosmo}-\eqref{KasnerCosmoExp}.  If the parameters of our solution are such that $\beta^2>1/2$, (corresponding to decreasing $g_{tt}$) the system will remain in that Kasner epoch all the way to  the final singularity at $z\to\infty$.
However, if $\beta^2<1/2$ (increasing $g_{tt}$) at the beginning of the Kasner epoch,  the evolution of the system is  more elaborate.
In this case, the growth of the Maxwell field \eqref{MaxKasner}  triggers a  transition to a new Kasner epoch with
$\beta\to \beta_\mathrm{new}$. 
When $\beta_{\rm new}^2 > 1/2$, the Kasner transition is often well described by  \eqref{eq:3rdODE} (as  demonstrated in Fig.~\ref{fig.KasnerInversions}). In such cases, the Kasner exponent changes sign 
 and the system then remains in this Kasner epoch with $g_{tt}$ decreasing toward zero at the  $z\to \infty$ singularity.

\begin{figure}[h!]
    \centering
    %\vspace{1cm}
    \includegraphics[width=1.0\textwidth]{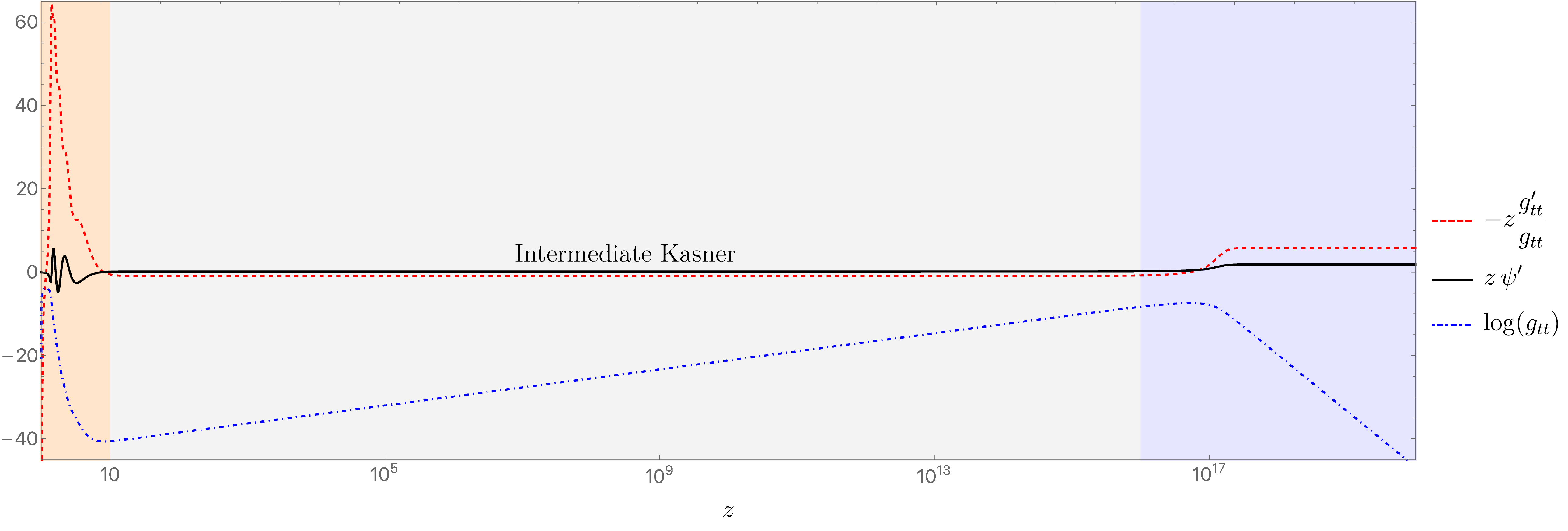}
    \caption{Typical example of the full dynamical evolution of a system with a Kasner transition like the ones observed in Fig.~\ref{fig.KasnerInversions}.  The system evolves from the event horizon at $z=1$  towards the final singularity at $z\rightarrow \infty$. In $1<z < 10$ (yellow shaded area) the system goes quickly through the ER bridge collapse and Josephson
    oscillations epoch, then it stays relatively steady for a long time $z$ (but short proper time) in the intermediate Kasner regime with $\beta\sim .1721212$, and finally (blue shaded area) there is a Kasner transition to $\beta_{\rm {new}}\sim 1.848235$.  This is for $\alpha = 1, q =m, mM =0.5,Q/Q_c-1=5.861297\times 10^{-5}$.}
       \label{fig.FullEvolution}
\end{figure}

An example of the complete dynamics including a Kasner transition is shown in 
Fig.~\ref{fig.FullEvolution}. The yellow shaded region shows the Josephson oscillations of the scalar field (solid black line) and its effect on $g_{tt}$ (blue dot-dashed curve). The series of short steps in $z\,g_{tt}^\prime/g_{tt}$, similar to the right plot of  Fig.~\ref{fig.josephson}, is also visible  (red dashed curve). The grey shaded region shows the intermediate Kasner regime with $\beta^2 < 1/2$. Since $g_{tt}$ is a power law, $z\,g_{tt}^\prime/g_{tt}$ is constant, and since $\beta^2 < 1/2$, $g_{tt}$ grows during this period. But eventually there is a transition to a new Kasner epoch with $\beta^2 > 1/2$ (blue shaded region) which continues until the singularity at $z = \infty$. Despite the fact that the intermediate Kasner region lasts for an exponentially large range of $z$, it corresponds to a short proper time.

The above discussion of Kasner transitions breaks down when $z\psi'$ becomes very large. This is because some of the terms we dropped in deriving \eqref{eq:3rdODE} are no longer negligible. This can be seen very easily in the AdS case with $\alpha =0$, since the inversion formula  $\beta_{\rm new} = 1/(2\beta)$ predicts $\beta_{\rm new}\to \infty$ when $\beta \to 0$. But an infinite $\beta_{\rm new}$ would correspond to a solution with a smooth Cauchy horizon, which is forbidden. In that case the terms in the interior equations of motion \eqref{eominterior} involving the charge $\tilde{q}$ of the scalar field can no longer be neglected \cite{Hartnoll:2020fhc}. They modify the 
 $\beta \rightarrow \beta_{\rm new}$ transition formula so that $\beta_{\rm new}$ remains finite. We find a similar breakdown to \eqref{eq:3rdODE}  when $\alpha > 0$. The main difference is that since the Kasner transition is no longer given by a simple inversion $\beta_{\rm new} = 1/(2\beta)$, the values of $\beta$ initially that lead to large $\beta_{\rm new}$ must be found numerically.

\begin{figure}[h!]
    \centering
    %\vspace{1cm}
    \includegraphics[width=0.95\textwidth]{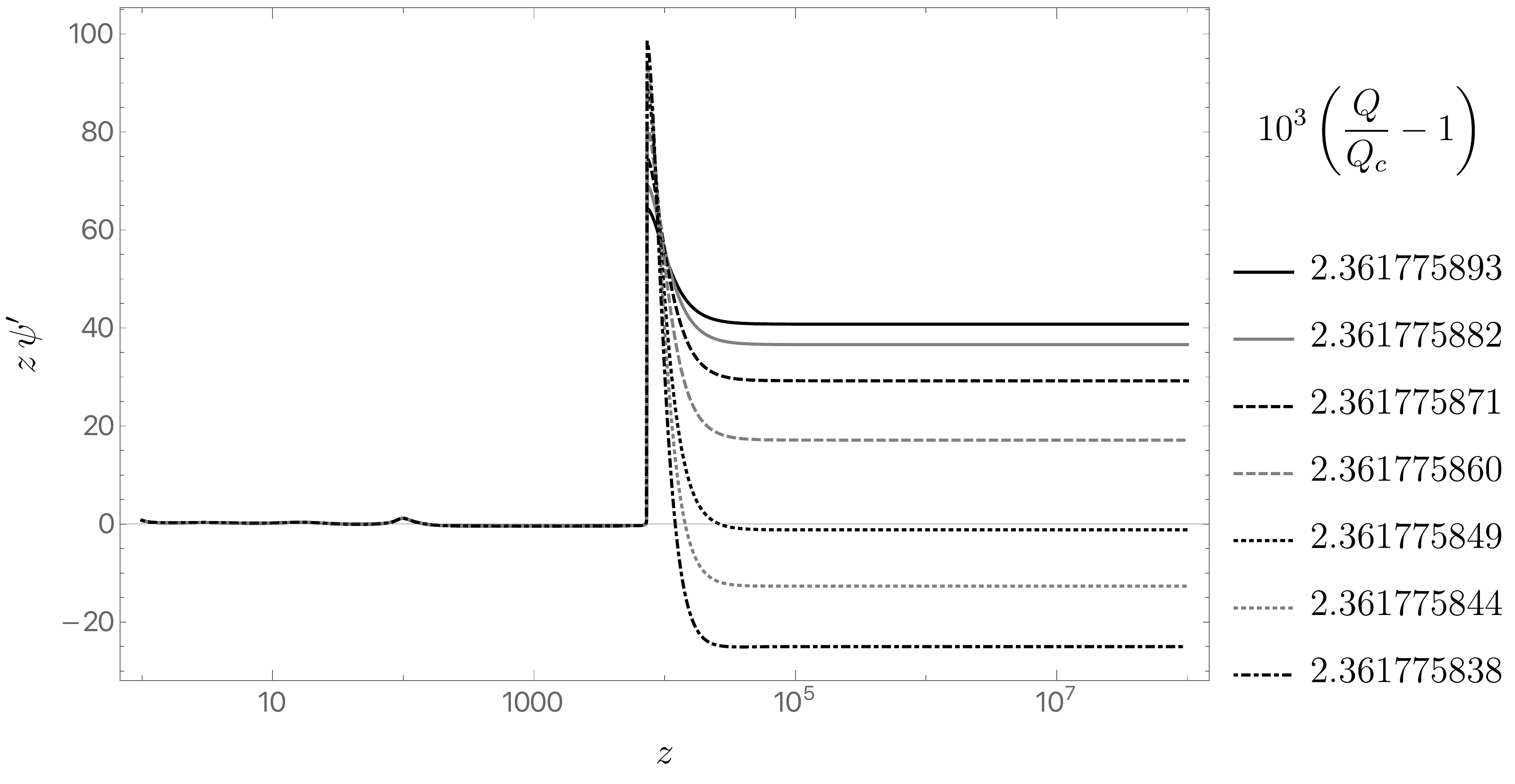}
    \caption{Examples of Kasner transitions that are not described by \eqref{def:S}. This is for $\alpha = 1, q = 0.5\,m, mM = 1.0$ and  several values of $Q/Q_c$ (see legend caption)  In all cases, the initial constant value of $\beta = z \psi' $ before the transition is close to $-0.42$. Note the high sensitivity of the curves to a small change of $Q/Q_c$.}
       \label{fig.KasnerInvAll}
\end{figure}

Examples of these more general Kasner transitions are presented in Fig.~\ref{fig.KasnerInvAll} for  $\alpha = 1, q = 0.5 m, mM = 0.5$ and slightly different values of $Q/Q_c$. Rather than having $z\,\psi^\prime$ grow monotonically as in  Fig.~\ref{fig.KasnerInversions}, we see that it now reaches a maximum before settling into $\beta_\mathrm{new}$. Remarkably, the new value of $\beta$ changes by multiples of ten when $Q/Q_c -1$ changes by $10^{-10}$! This highlights the extreme sensitivity of the interior dynamics on the black hole charge.

\begin{figure}[h!]
    \centering
    %\vspace{1cm}
    \includegraphics[width=0.95\textwidth]{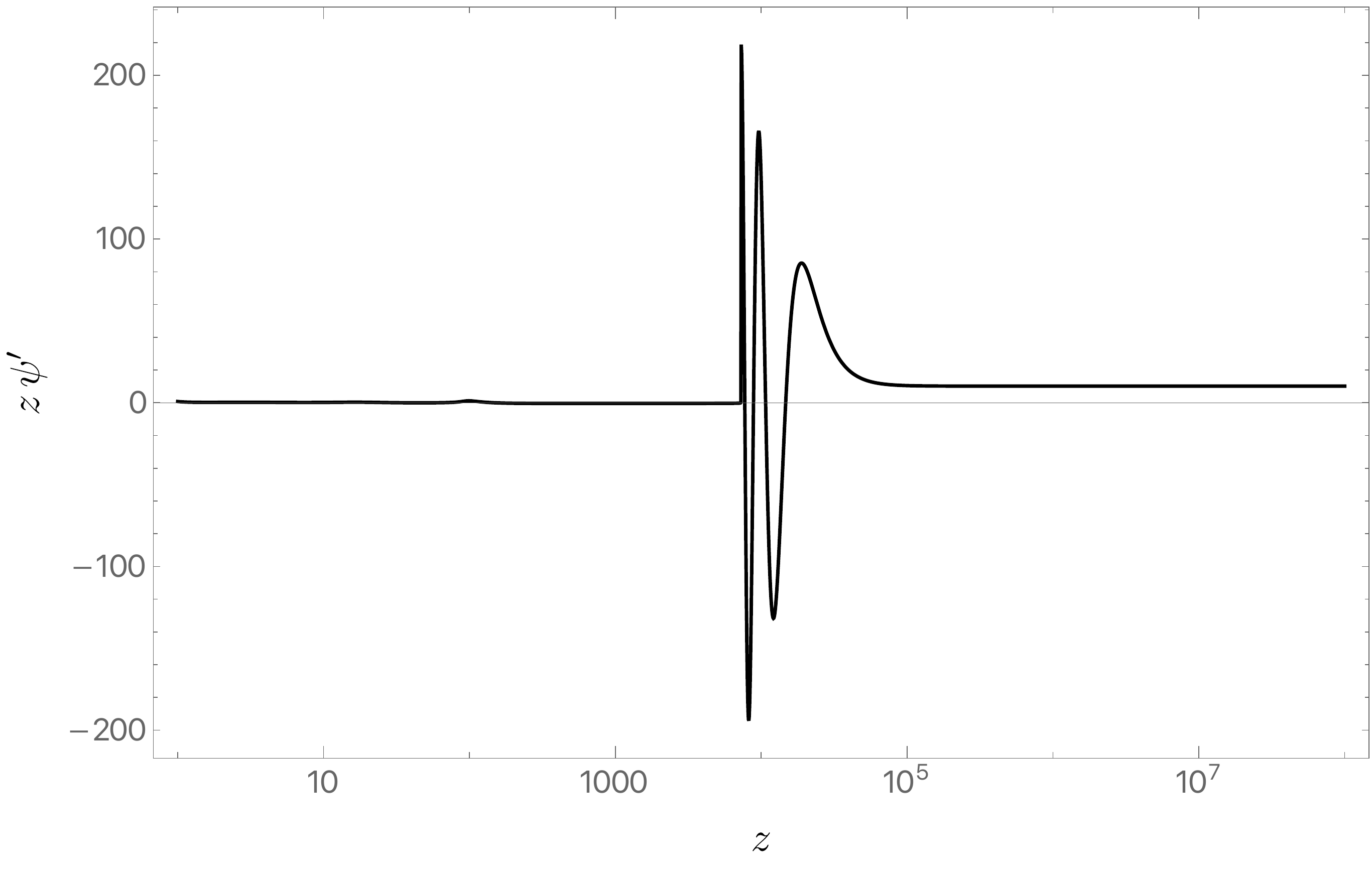}
    \caption{Details of a Kasner transition that is not described by \eqref{eq:3rdODE} and where a fine tuned choice of parameters ($\alpha = 1, q = 0.5\,m, mM = 1.0$ and $Q/Q_c-1= 2.361775783\times 10^{-3}$) shows that we can have Kasner transitions that are immediately followed by a `second' Josephson  oscillations phase. One has $\beta=-0.427266$ and $\beta_\mathrm{new}=1.01622$.}
       \label{fig.KasnerInvOsc}
\end{figure}

When the Kasner transition becomes sharp enough, it can trigger a new round of Josephson oscillations, as shown in
Fig.~\ref{fig.KasnerInvOsc}. Note that this figure has the same parameters as Fig.~\ref{fig.KasnerInvAll}, with only a tiny change in the charge. The presence of these oscillations clearly shows the importance of the charge terms that were dropped in \eqref{eq:3rdODE}. Unlike the original Josephson oscillations, these later oscillations can also depend on $\alpha$.

\begin{figure}[h!]
    \centering
    %\vspace{1cm}
    \includegraphics[width=0.95\textwidth]{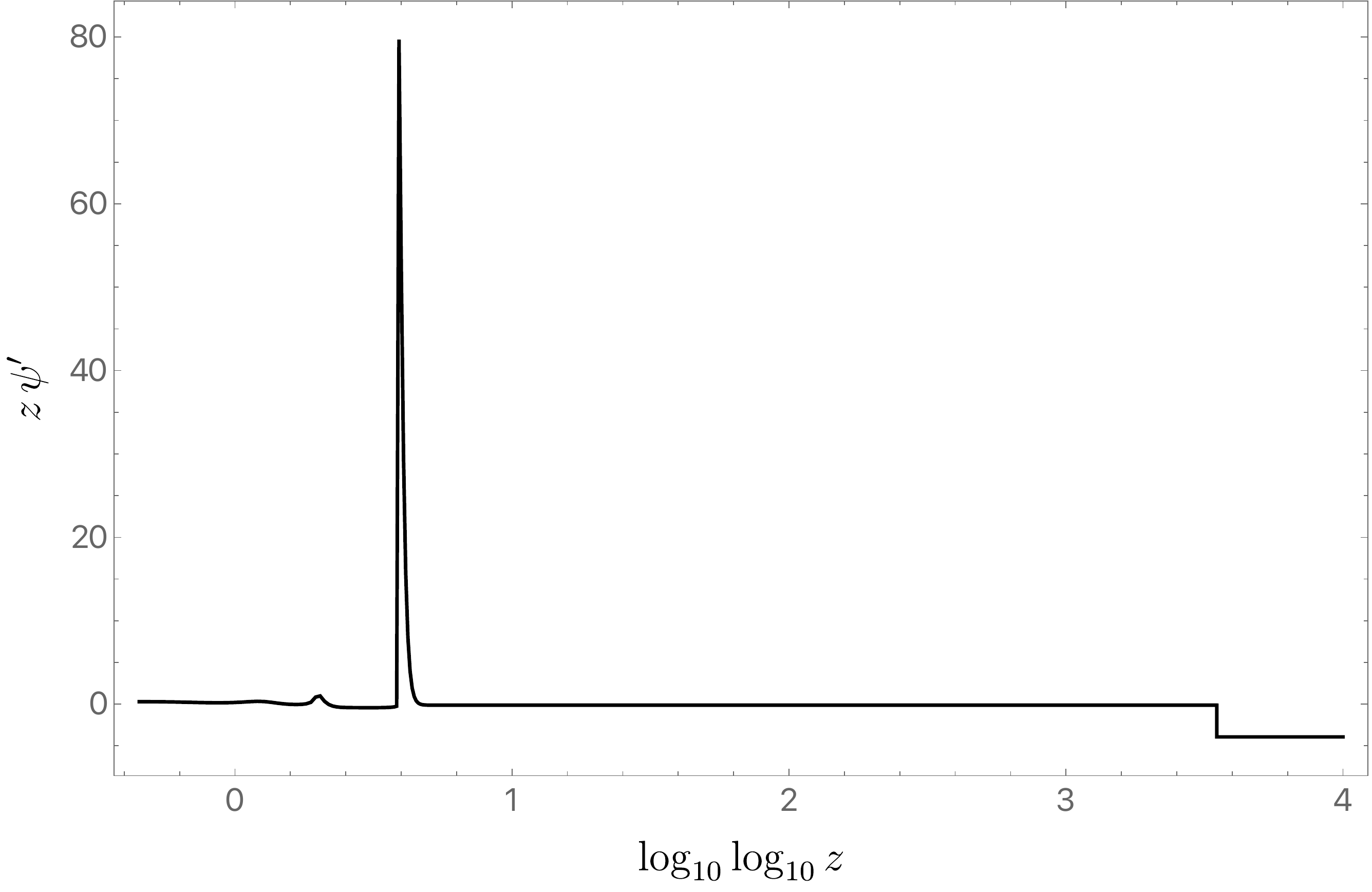}
    \caption{Example of multiple Kasner transitions which occur when the first transition yields $|\beta_\mathrm{new}|<1/\sqrt{2}$.
    For $\alpha = 1, q = 0.5\,m, mM = 0.5$ and $Q/Q_c-1= 2.361775850\times 10^{-3}$, the system starts with $\beta=-0.426086$, and it 
    has  $\beta_\mathrm{new}=-0.126666$ after the first transition which triggers a second transition at a very large $z$ of $z_\mathrm{tr \,(2)}\sim 3.3017\times 10^{3507}$ (note the double logarithmic scale) that produces $\beta_{\mathrm{new}\,(2)}=-3.93114$.}
       \label{fig.KasnerInvSecond}
\end{figure}

 Fig.~\ref{fig.KasnerInvAll} shows that $\beta_{\rm new}$ can be both positive and negative depending on the charge. Since we can vary $Q$ continuously, there are clearly cases with  $\beta_{\rm new}^2 < 1/2$. 
 The subsequent evolution  will then lead to a second Kasner transition similar to the first. 
This is illustrated in Fig.~\ref{fig.KasnerInvSecond} where we show a case where a second Kasner transition indeed occurs at an incredibly large $z$. 
For certain finely tuned values of $Q$, this will repeat an arbitrarily large number of times.

\section{Discussion}

We have studied the dynamics inside a class of asymptotically flat hairy black holes. The evolution turns out to resemble the intricate structure recently found inside a holographic superconductor \cite{Hartnoll:2020fhc}. Just above the critical charge $Q_c$ where the Reissner-Nordstr\"om solution becomes unstable to forming scalar hair, the interior dynamics passes through three epochs. There is a collapse of the Einstein-Rosen bridge, Josephson oscillations of the scalar field, and then a Kasner epoch sometimes followed by transitions to different Kasner epochs. The Cauchy horizon is gone, and the spacetime ends in a Kasner singularity. But the parameter $\beta$ governing the Kasner singularity \eqref{KasnerCosmoExp} depends very sensitively on the black hole charge, as shown in  Fig.~\ref{fig.osc}. As $Q\rightarrow Q_c $ from above, and the scalar field outside the horizon smoothly goes to zero, the Kasner epoch inside cycles through a finite range of parameters an infinite number of times.

In addition, whenever $\beta^2 < 1/2$, the evolution for fixed $Q$ involves a  transition to a new Kasner epoch. This transition can trigger very fine oscillations resulting in the new epoch  again  having $\beta^2 < 1/2$, leading to further transitions. The number of such transitions depends very sensitively on $Q$ and can be arbitrarily large.  The net result is a chaotic pattern of singularities inside black holes with charge near $Q_c$. This is reminiscent of the mixmaster behavior expected near a generic singularity \cite{Damour:2002et}, but the mechanism is different. In the present case, the charged scalar field plays a crucial role.

\begin{figure}[h!]
    \centering
    %\vspace{1cm}
    \includegraphics[width=0.95\textwidth]{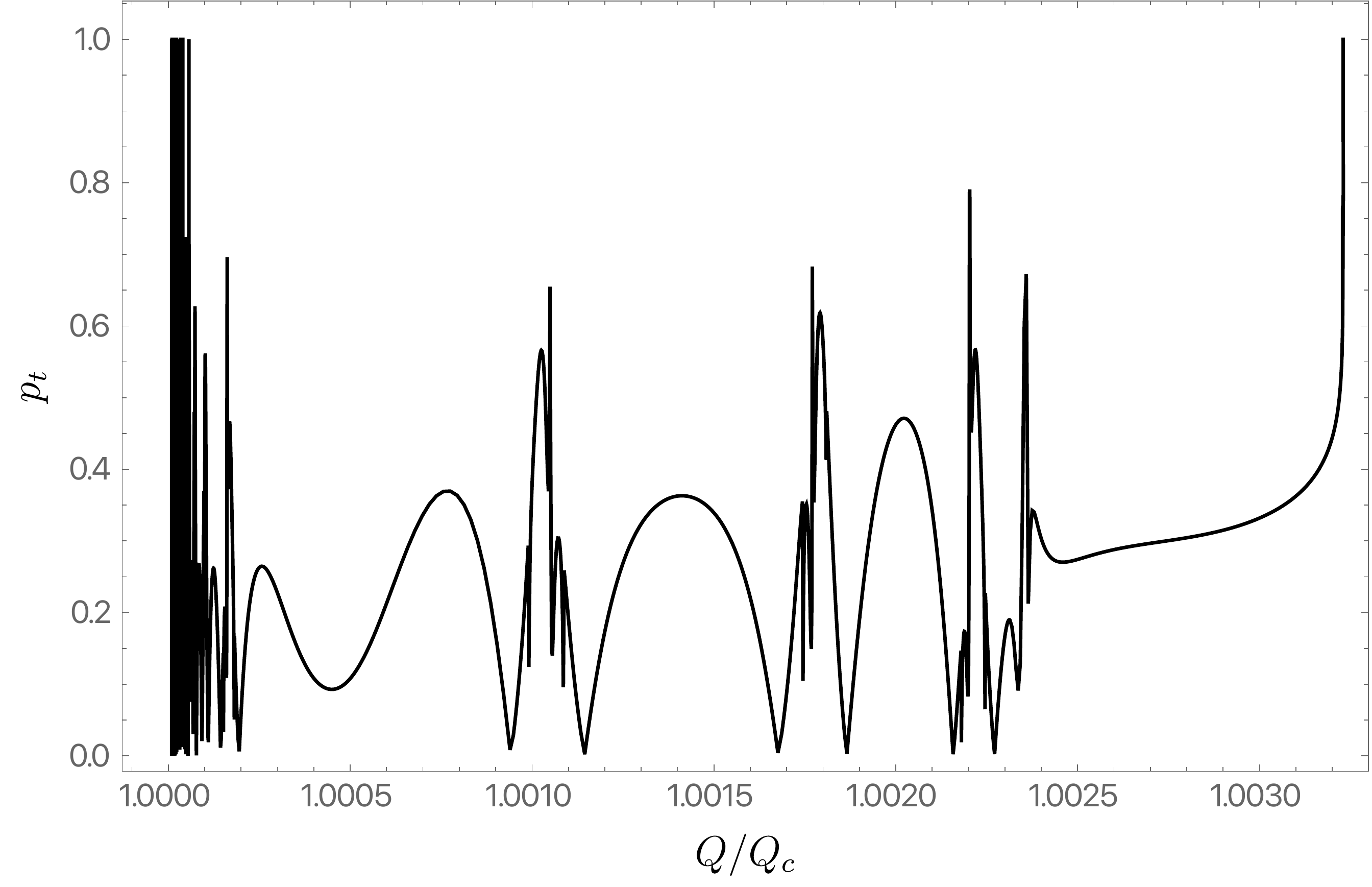}
    \caption{The Kasner exponent $p_t$ as a function of $Q/Q_c$ from the onset of the scalar instability to the maximum charge this black hole can carry. This is for $\alpha = 1, \ q = .5m$ and $mM = 1$. The solution with maximum $Q/Q_c$ is singular.}
       \label{fig.globalpt}
\end{figure}

We have focussed on the case where the charge is slightly above $Q_c$ since that is where the dynamics cleanly separates into these different epochs, and the singularity exhibits chaotic behavior. A more global picture of the singularity inside these hairy black holes is given in Fig.~\ref{fig.globalpt}, which shows the Kasner exponent $p_t$ over the entire range from $Q=Q_c$ to the maximum charge that the black hole can carry.

As mentioned in the introduction, it was shown in \cite{Dias:2021vve} that these hairy black holes can have unusual extremal limits. For a certain range of parameters, the maximum charge black hole for given mass is a nonsingular black hole with  $Q^2>M^2$ and nonzero temperature. This is possible since the scalar field is only bound to the black hole if $\mu$ (the electrostatic potential difference between the horizon and infinity) satisfies $ \mu^2 q^2 \le m^2$. This bound can be saturated when the black hole still has $T>0$. The resulting objects were called ``maximal warm holes". 

\begin{figure}[h!]
    \centering
    %\vspace{1cm}
    \includegraphics[width=0.95\textwidth]{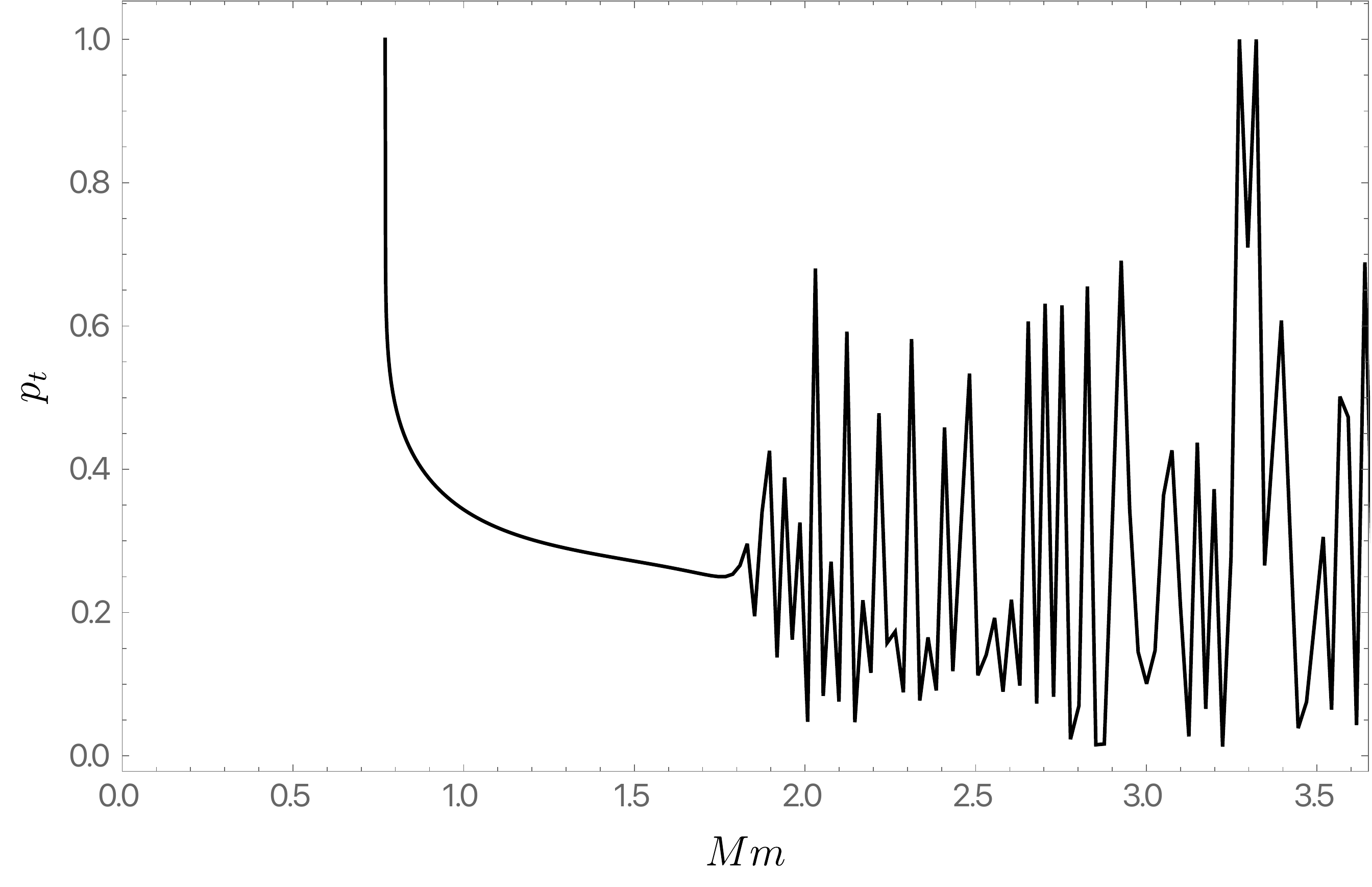}
    \caption{The Kasner exponents inside a family of maximal warm holes (the red curve family of Fig.~1 in \cite{Dias:2021vve}). These are maximally charged black holes with $\alpha = 1, \ q=m$. The wiggles on the right are a result of the solution having a charge that approaches $Q_c$. At the lower limit of the mass, the solution becomes singular.}
       \label{fig.warmholes}
\end{figure}

We have examined the singularity inside these maximal warm holes. The Kasner exponents are shown in Fig.~\ref{fig.warmholes} {for the maximal warm hole family described by the red curve in the phase diagram of Fig.~1 of \cite{Dias:2021vve}}. As $Mm$ increases, the difference between the maximum charge, {\emph{i.e.} maximal warm hole charge}, and critical charge for the {instability of} a {RN} black hole of that mass decreases to zero. So the wiggles seen on the right side of this figure can be understood as a result of the maximal warm hole approaching the onset of the {RN} instability. In both this figure and the previous one, $p_t \rightarrow 1$ as the solution approaches a singular extremal solution. This is because the Kasner singularity approaches the event horizon. In other words, the proper time from the bifurcation surface to the singularity goes to zero as one approaches the singular extremal solution.

 Although we have not studied fully nonlinear dynamical solutions in our theory, it is very plausible that if we start with generic initial data for RN in its unstable regime, it will evolve to one of our static hairy black holes. The fact that these black holes do not have a Cauchy horizon is some evidence that strong cosmic censorship holds in this theory.

Our discussion has been entirely classical, and we expect quantum effects to become important near the singularity. They will also modify the Josephson oscillations when $Q\rightarrow Q_c$, and the frequency of oscillations diverge. However,  staying away from the Planck regimes, we find it remarkable that adding a small amount of scalar hair can change a black hole interior in such profound and sensitive ways.

We have focused on static charged black holes, but Kerr black holes can also support massive scalar hair \cite{Herdeiro:2014goa,Chodosh:2015oma,Chodosh:2015nma,Herdeiro:2015gia}. In fact, these are solutions which are not stationary and axisymmetric, and have only a single Killing field \cite{Dias:2011at,Herdeiro:2014goa,Chodosh:2015oma,Chodosh:2015nma,Herdeiro:2015gia,Dias:2015rxy}. A complete investigation of their properties inside the horizon is beyond the scope of this paper, but in the Appendix we show that they cannot have a smooth Cauchy horizon. The proof is quite general and applies to black holes in more than four dimensions also.

\subsection*{Acknowledgments}
O.~J.~C.~D. acknowledges financial support from the STFC Grants ST/P000711/1 and ST/T000775/1.
The work of G.~T.~H. was supported in part by NSF Grant PHY-2107939. J.~E.~S has been partially supported by STFC consolidated grants ST/P000681/1, ST/T000694/1.
The numerical component of this study was partially carried out using the computational facilities of the Fawcett High Performance Computing system at the Faculty of Mathematics, University of Cambridge, funded by STFC consolidated grants ST/P000681/1, ST/T000694/1 and ST/P000673/1. The authors further acknowledge the use of the IRIDIS High Performance Computing Facility, and associated support services at the University of Southampton, in the completion of this work.

\appendix

%%%%%%%%%%%%%%%%%%
\section{No smooth Cauchy horizon for rotating  black holes with scalar hair}
%%%%%%%%%%%%%%%%%%

We consider gravity in $d$-dimensions coupled to a complex scalar field with action 
\begin{equation}
S=\frac{1}{16\pi G}\int \mathrm{d}^d x\sqrt{-g}\left[R-\nabla_a \phi\nabla^a \phi^\star -V(\phi \phi^\star)\right]\,,
\end{equation}
where ${}^*$ denotes complex conjugation. The equations of motion are
\begin{subequations}
\begin{align}\label{eq:eom}
& R_{ab}= \frac{1}{2}\nabla_a \phi \nabla_b \phi^\star+\frac{1}{2}\nabla_b \phi \nabla_a \phi^\star+\frac{V}{d-2}g_{ab}
\\
& \Box \phi-V^\prime(\phi \phi^\star) \phi=0\,.
\end{align}
\end{subequations}
We are interested in solutions where the metric is stationary and axisymmetric. As such, the metric admits two commuting Killing fields $k$ and $m$.  We introduce two coordinates $t\in\mathbb{R}$ and $\varphi\in[0,2\pi)$, so that $k=\partial/\partial t$ and $m=\partial/\partial_\varphi$.
We assume the scalar field is nonzero and not invariant under $k$ nor $m$, but only a linear combination.  We also assume the solution has a smooth Cauchy horizon and will derive a contradiction.

Since we are interested in asymptotically flat black holes, the intersection of a partial Cauchy surface of constant $t$ and the horizon (either the inner horizon or outer) must be compact. Under these circumstances, we expect  that the horizons are Killing horizons with horizon generators $\xi^{\pm}$.

We can use the symmetries of the metric to write
$$
\phi = e^{-i\omega t+i\,m_{\varphi} \varphi}\hat{\phi}
$$
where $k(\hat{\phi})=m(\hat{\phi})=0$, and $m_{\varphi}\in\mathbb{Z}$. The rigidity theorem tells us that if a Killing vector field exists, and is not the horizon generator, then a second Killing field must exist. Since we want to evade this theorem, we take the scalar to be invariant under the black hole horizon generator $\xi^+=k+\Omega^+ m$  only ($\Omega^+$ is the angular velocity of the outer horizon). This implies we must take
$$
\omega = m_{\varphi} \Omega^+\,.
$$
Applying the same reasoning to the Cauchy horizon requires ($\Omega^-$ is the associated angular velocity)
$$
\omega = m_{\varphi} \Omega^-\,,
$$
from which we conclude that the scalar $\phi$ will only be invariant and regular at the Cauchy and black hole event horizons if $\Omega_+=\Omega_-$, which implies $\xi^+=\xi^-\equiv \xi$.

Since $\xi$ is Killing we have
\begin{equation}\label{KillingProperties}
\nabla_a \xi_b+\nabla_b\xi_a=0\quad\text{and}\quad {\xi}^a \nabla_a \phi=0\,.
\end{equation}
Using the Ricci identity for co-vectors $X$
\begin{equation}
\nabla_a \nabla_b X_c-\nabla_b\nabla_a X_c=R_{abcd}X^d\,,
\end{equation}
we  have
\begin{equation}
 \nabla^b\nabla_b \xi_a=-\nabla^b\nabla_a\xi_b=-R^{b}_{\phantom{b}abc}\xi^c-\nabla_a \nabla_b\xi^b=-R_{ab}\xi^b
 %\Rightarrow \Box \xi_a+R_{ab}\xi^b=0\,,
\end{equation}
where in the first equality we used \eqref{KillingProperties} which also implies that $\nabla_b \xi^b=0$ for Killing fields. We use the latter to obtain the last equality.

In differential form notation this can be written\footnote{We use the differential forms' conventions of \cite{Dias:2019wof}.}
\begin{equation}
(\star \mathrm{d}\star \mathrm{d}\xi)_a = 2 R_{ab}\xi^b\equiv K_a
\end{equation}
or
\begin{equation}
\mathrm{d}\star \mathrm{d}\xi=\star K\,.
\end{equation}
 Let $\Sigma^+_-$ be a timelike surface of constant $t$ that extends from the  Cauchy horizon to the black hole event horizon. Note that $\star K$ is a $(d-1)$-form so it makes sense to integrate it over $\Sigma^+_-$. It then follows that
\begin{equation}\label{keyRelation}
\int_{\Sigma^+_- }\star K=\int_{\Sigma^+_- } \mathrm{d}\star \mathrm{d}\xi=\int_{\partial \Sigma^+_- } \star \mathrm{d}\xi\,.
\end{equation}
By standard arguments (see \emph{e.g.} section 12.5 of \cite{Wald:1984rg}), the last  term in \eqref{keyRelation} is simply $-16\pi G\,(T_+ S_++T_- S_-)$, and is manifestly negative.  The first term can be simplified using  the field equation \eqref{eq:eom}
and the last relation in \eqref{KillingProperties} to obtain
\begin{equation}
K_a = 2 R_{ab}\xi^b = \frac{2V}{d-2} \xi_a\,.
\end{equation}
Altogether, \eqref{keyRelation} finally reads
\begin{equation}
T_+ S_+ + T_- S_- =-\frac{1}{8\pi G\,(d-2)}\int_{\Sigma^+_-} V \xi^a n_a \sqrt{-h}\,\mathrm{d}^{d-1}x,
\end{equation}
where $h$ is the induced metric on $\Sigma^+_-$ and $n^a$ is its spacelike unit normal (directed toward increasing $t$). However, if we
note that $\xi^a n_a = n_t=1/\sqrt{g^{tt}}>0$,
we see that for any positive potential such as a simple mass term
 $V=m^2 |\phi|^2$, the right hand side is negative definite. So we have reached a contradiction.
This establishes that stationary hairy black holes of 
Einstein's gravity coupled to a complex scalar with positive potential do not have smooth Cauchy horizons.

\bibliographystyle{jhep}
	\cleardoublepage

\renewcommand*{\bibname}{References}

\bibliography{kasner}

\end{document}